\documentclass[aps,pre,twocolumn,superscriptaddress,nofootinbib]{revtex4}

\usepackage{amssymb,amsfonts,amsmath,graphicx,stmaryrd}

\newcommand{\IO}{input-output }

\newcommand{\beq}{\begin{equation}}
\newcommand{\eeq}{\end{equation}}
\newcommand{\beqn}{\begin{eqnarray}}
\newcommand{\eeqn}{\end{eqnarray}}
\renewcommand{\th}{\vartheta}

\newcommand{\suppress}[1]{}

\renewcommand{\d}[2]{\frac{d #1}{d #2}}
\newcommand{\pd}[2]{\frac{\partial #1}{\partial #2}}

\begin{document}

\title{A statistical method for revealing form-function relations in biological networks}

\author{Andrew Mugler\footnote{Current address: FOM Institute for Atomic and Molecular Physics (AMOLF), Science Park 104, 1098 XG Amsterdam; {\tt mugler@amolf.nl}}}
\affiliation{Department of Physics, Columbia University, New York, NY 10027}

\author{Boris Grinshpun}
\email[]{bg2178@columbia.edu}
\affiliation{Department of Applied Physics and Applied Mathematics, Columbia University, New York, NY 10027}

\author{Riley Franks}
\affiliation{Department of Applied and Computational Mathematics, California Institute of Technology, Pasadena, CA 91125}

\author{Chris H. Wiggins}
\affiliation{Department of Applied Physics and Applied Mathematics, Center for Computational Biology and Bioinformatics, Columbia University, New York, NY 10027}

\begin{abstract}
Over the past decade, a number of researchers in systems biology 
have sought to relate the function of biological systems 
to their network-level descriptions
--- lists of the most important players and the pairwise interactions 
between them. Both for large networks (in which statistical 
analysis is often framed in terms of the abundance of repeated 
small subgraphs) and for small networks which can be analyzed in 
greater detail (or even synthesized in vivo and subjected to 
experiment), revealing the relationship between the topology of 
small subgraphs and their biological function has been a 
central goal. We here seek to pose this revelation as a 
statistical task, illustrated using a particular
setup which has been constructed experimentally and for which 
parameterized models of transcriptional regulation have been 
studied extensively.
The question ``how does function follow form" is here mathematized
by identifying which topological attributes correlate with the 
diverse possible information-processing tasks which a 
transcriptional regulatory network can realize. 
The resulting method reveals 
one  form-function relationship which had earlier been predicted 
based on analytic results, and reveals a second for which we can 
provide an analytic interpretation. Resulting source code is 
distributed via {\tt http://formfunction.sourceforge.net}.
\end{abstract}

\maketitle

The observation that form constrains function in biological 
systems has historical roots far older than
systems biology. Century-old 
examples include those made in D'Arcy Thompson's ``On Growth and Form" 
\cite{thompson1917form} and the observation that the quick responses 
necessary for reflex actions such as heat- and pain-avoidance 
could be manifest only by a dedicated input-output relay circuit 
from fingers to brain and back \cite{sherrington}. 
Advances in synthetic biology, often requiring design of systems
for which  only topology can be specified without control over precise
parameter values, has motivated a reintroduction of such topological
thinking in biological systems \cite{Stricker, Gardner, Elowitz}.
A second source of such inquiry
is high-throughput systems biology, in which technological advances
provide topologies of large biological networks without precise 
knowledge of their interactions, dynamics, or possible naturally-occuring inputs
\cite{Cusick, Schena, Giot}.
Such limitations thwart our desire to learn form-function 
relations from data or to 
derive them from plausible first-principles modeling.
Our goal here is to illustrate how re-framing 
the question as one of computation and statistical analysis 
allows a clear, quantitative, interpretable approach.

\section{Setup}

\begin{figure}
\begin{center}
\includegraphics[scale=0.63]{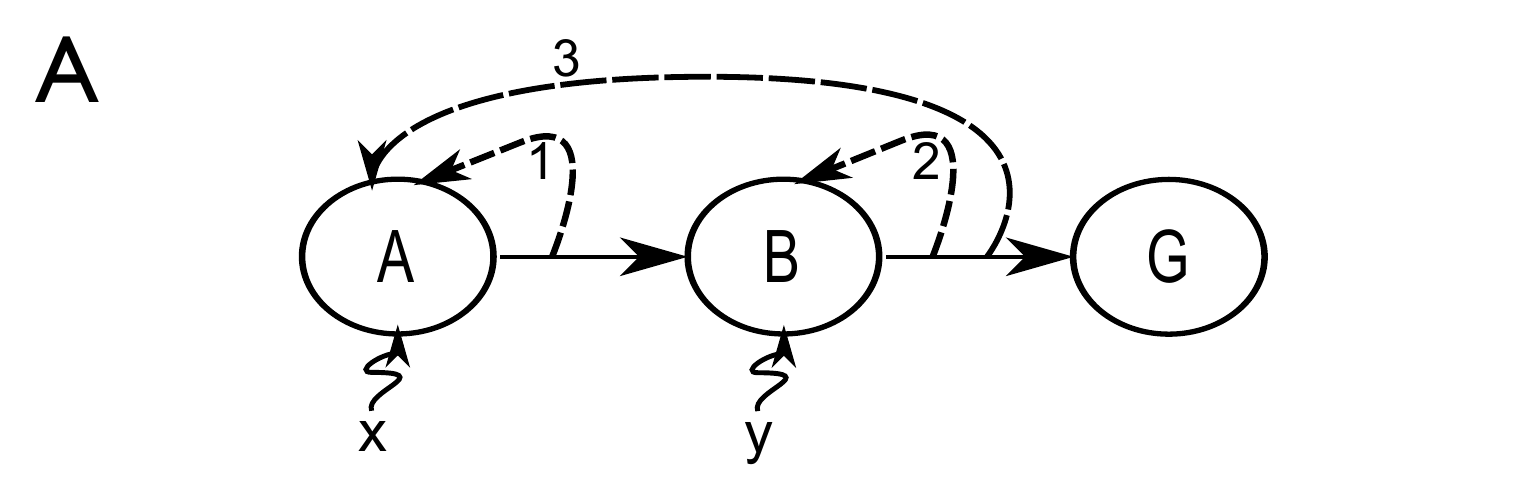}
\includegraphics[scale=0.21]{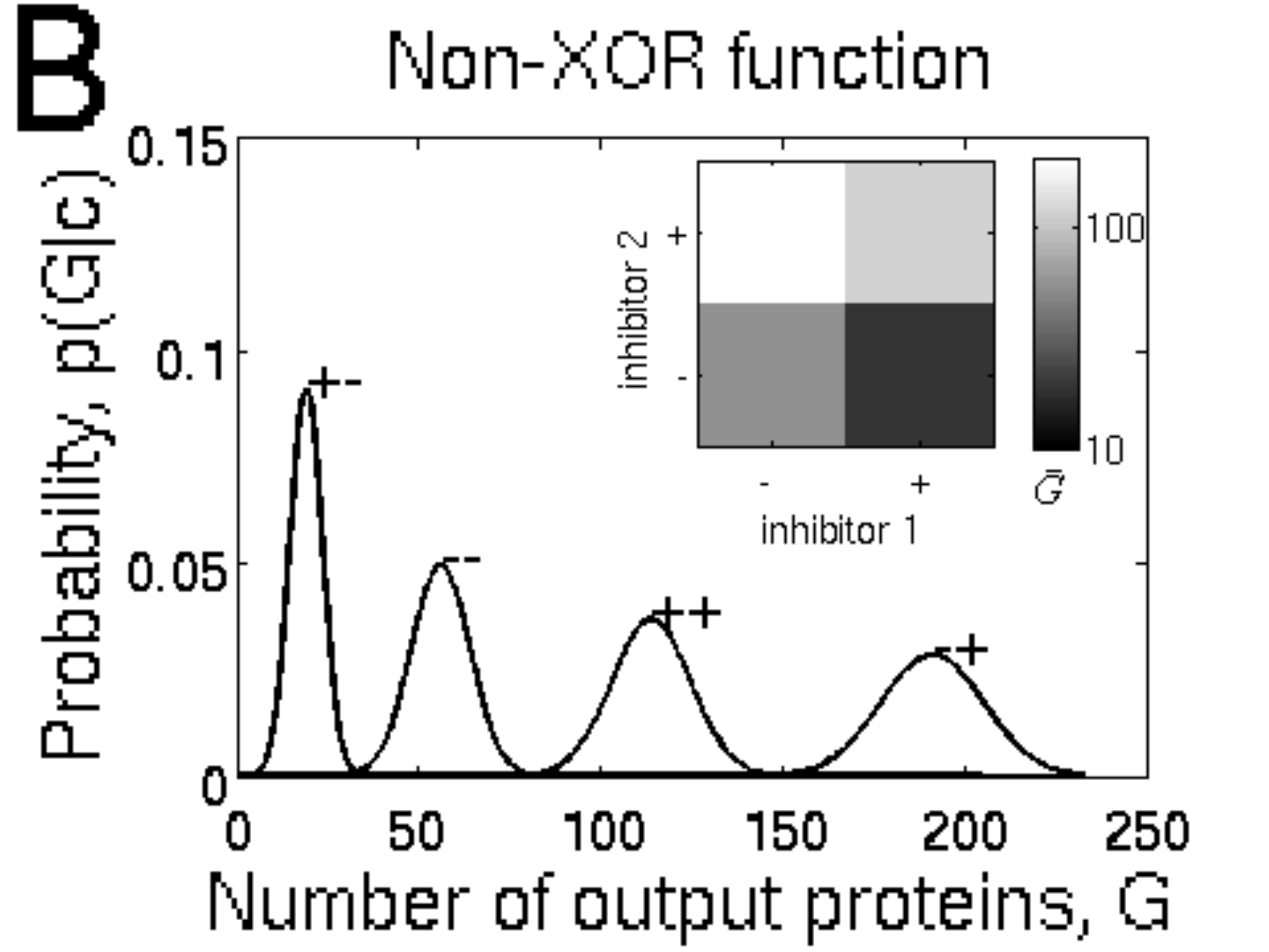}
\includegraphics[scale=0.21]{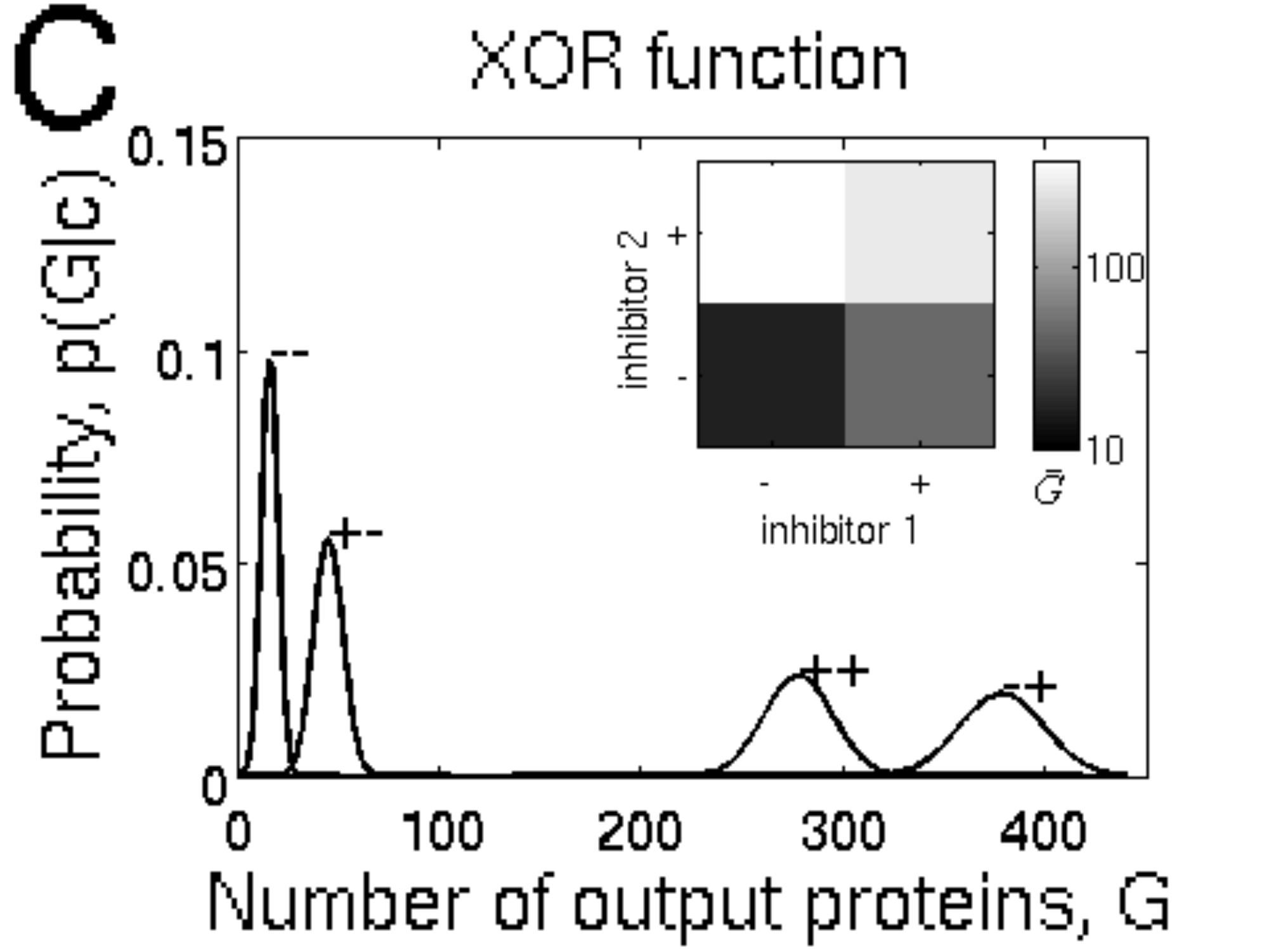}
\caption{Network set and input-output functions.  (A) Transcription factor $A$ regulates the expression of transcription factor $B$, which regulates the expression of fluorescent protein $G$.  The efficacies of $A$ and $B$ are reduced by the presence of chemical inhibitors, labeled by scaling factors $x$ and $y$ (Eqn.\ \ref{eq:dynsys}), respectively.  We distinguish between up- and down-regulation and consider all ways in which regulatory edges $1$, $2$, and $3$ may appear,
for a total of $160$ networks.  (B-C) Examples of non-XOR (B) and XOR (C) functions (see Sec.\ \ref{sec:nonpar}), as defined by the ranking of conditional probability distributions $p(G|c)$, where $c \in \{ --, -+, +-, ++ \}$ describes whether each inhibitor is present ($+$) or absent ($-$).  Insets show mean protein number $\bar{G}$ in each of the four states.  The functions in B and C are both realized by the particular network in A in which edge 3 is absent and all remaining edges are down-regulating.}
\label{fig:cartoon}
\end{center}
\end{figure}

Mathematical progress requires clear definitions of terms, 
including, here, ``form," ``function," and ``follow." To define 
the first two we must choose a specific experimental setup; we here choose 
one which has been experimentally realized repeatedly: that of a 
small, synthetic transcriptional regulatory network with 
``inducible promoters," meaning that the efficacy of the 
transcription factors may be diminished by introducing small 
interfering molecules \cite{Guet,Elowitz,Basu,Mangan}
(Figure \ref{fig:cartoon}A).
A common ``output" responding 
to the ``input" presence of such small molecules is
the abundance of inducible green fluorescent protein (GFP), which provides
an optical readout of one of the regulated genes. The ``form," then, 
will be defined by the topology of such a small regulatory network, 
distinguishing between up- and down-regulatory edges in the 
network.\footnote{We use $\rightarrowtriangle$ to indicate up-regulation, $\dashv$ to indicate down-regulation, and $\rightarrow$ to indicate regulation whose sign is not specified; additionally we use $\leadsto$ to indicate inhibition by a small molecule.}
``Function" will be defined by the realizable input-output relations of
a device with two binary inputs (corresponding to 
presence or absence of the small molecules) and one real-valued 
output (the transcriptional level of the output gene).

Among other published experiments which correspond to this setup 
is that of Guet, et al.\ \cite{Guet}. We remind the reader of two 
particularly noteworthy observations of Guet and coworkers: (i) 
that many of their experimentally-realized small networks were 
``broken" in the sense that the output remained constant over the 
different possible inputs; and (ii) that often the same topology 
can realize different input-output relations (or even be broken, 
i.e., can realize both a particular function as well as a lack 
of function entirely).
Within a mathematical model, such behavior 
follows straightforwardly from considering the behavior of a 
given dynamical system at different points in the space of 
quantitative parameters \cite{strogatz}. Revealing such 
degeneracy of functions by exploring the parameter space 
given a topology (and 
given an algebraic 
expression modeling the regulatory interactions among the genes) 
may be recast as one of optimizing --- locally in parameter space --- the mutual 
information between input and output over this space \cite{Ziv, Mugler1, Mugler2}.
Mutual information (MI) as a cost function
is advantageous both biologically (in that many natural systems
including transcriptional regulatory networks are known to 
operate near their information-optimal constraints 
\cite{Tkacik2, Mehta, WalczakII}) and mathematically (in that by 
optimizing MI we can identify parameter settings which are 
functional without demanding in advance the particular 
input-output functions we seek). That is, we optimize for 
functionality rather than for a specific predetermined
function.

MI between input and output is defined as a functional
of the joint distribution $p(c,G)$ --- the probability of a
(here, categorically-valued) setting of the chemistry and
the (here, real-valued) concentration of the output gene
(see Sec.\ \ref{sec:method}).
The stochastic relationship between input in output in biological networks
has many underlying sources; however, one source is intrinsic: the finite
copy number of the constituent proteins introduces a ``shot noise,"
much discussed in the systems biology literature \cite{Gillespie,Paulsson,Elowitz_noise,mtv,cspan}.
Particular additional sources of 
noise may thwart information processing as well in specific systems; hoping to 
remain as general as possible, we will here consider only intrinsic noise.
In this respect, we aim to describe the functional response(s)
of single cells, as opposed to an averaged response of a pooled population of cells.

Having defined form and function, we must define ``follow." Here 
we will need a measure of, for any topological feature of the 
networks, the extent to which the  value of the feature (e.g., 
length of a cycle, number of down-regulations, etc.) does or 
does not correlate with the input-output relations the network 
can perform. Since we wish to correlate two categorical (rather 
than real-valued) features, MI is again useful, in that we will 
rank topological features by the information between its value 
and the particular input-output relations networks with that 
feature are found to perform.

To summarize: ``function" is given by the possible input-output 
relations of which a given topology is capable; these are found 
by numerical optimization over parameters of the MI between 
input and output, with the probabilistic description of the 
transcriptional output set by intrinsic noise; ``form" is 
mathematized by enumerating a list of topological attributes 
descriptive of our small networks. The question ``how does 
function follow form" is here recast as a ranking of topological 
features based on the correlation (here, given by MI) between 
topological feature value and the particular \IO relations realizable for a given topology.

\section{Method}\label{sec:method}

\begin{figure}
\begin{center}
\includegraphics[scale=.35]{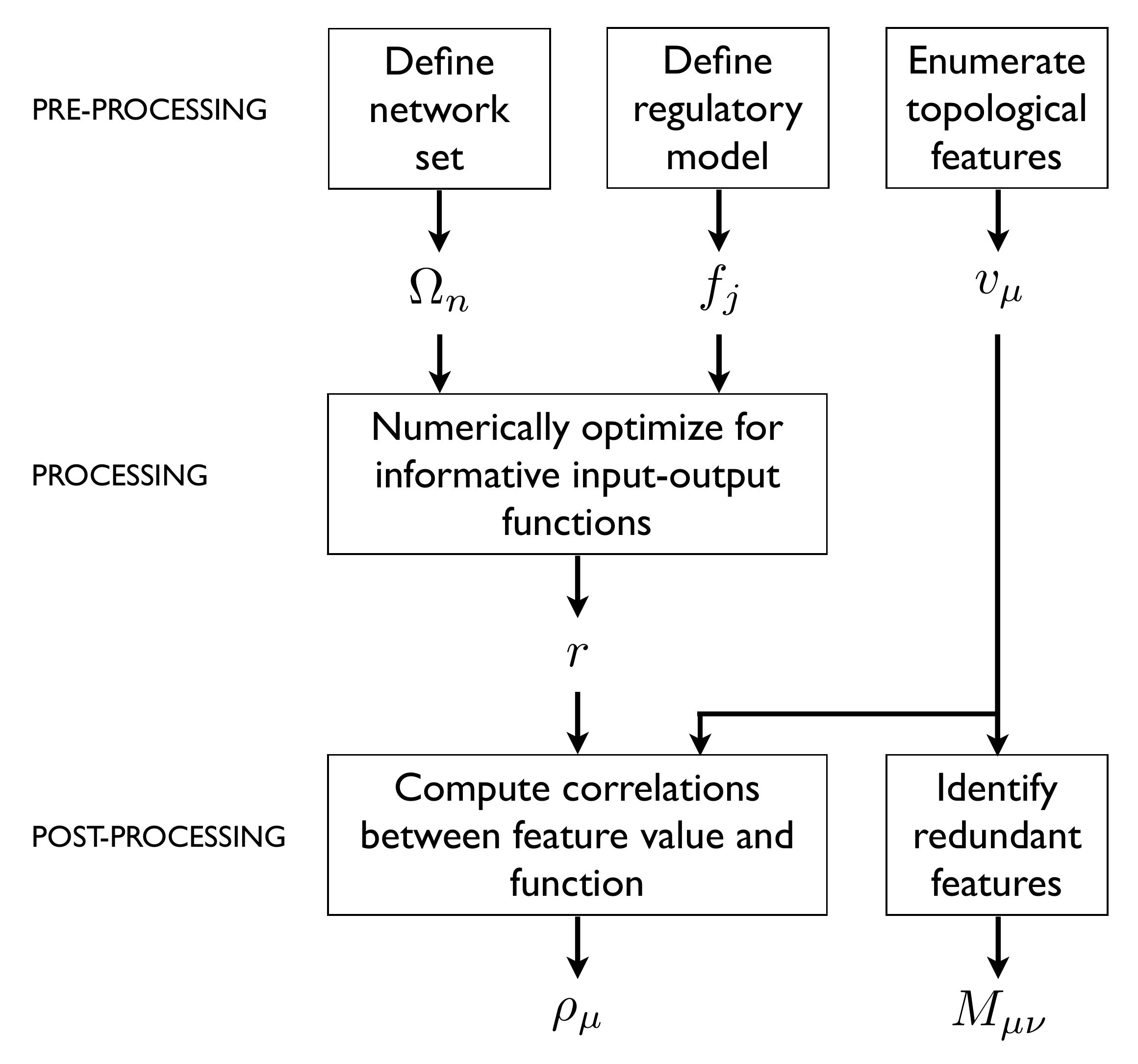}
\caption{Outline of the procedure used to ask ``how does function follow form.''
Variables are defined in text.}
\label{fig:pipeline}
\end{center}
\end{figure}

The method is illustrated in Fig.\ \ref{fig:pipeline}, which makes clear
how the (general) search for form-function relations may be composed into 
separate tasks, the implementation of which is particular to one particular
experimental setup. For example, the particular elements of the 
network set $\Omega_n$ follow from system-specific choices made below,
but the design of the task (Fig.\ \ref{fig:pipeline}) can be
applied to relating form and function more generally. For a different
experimental context, one or more of the ``modules" in the design of the
algorithm may need to be replaced, but the design itself we anticipate
will be useful to revealing form-function relations in a wide variety
of contexts.

Below we describe the method in detail as applied to our particular experimental setup; further information on networks, gene regulation, linear noise analysis, information theory, and optimization is provided in Sec.\ \ref{sec:app}.

\subsection{Networks and features}

We consider the simplest set of networks with two chemical inputs and one genetic output (Fig.\ \ref{fig:cartoon}A).  Each of two chemical inhibitors
is either present or absent, giving four possible input states.
When present, the chemicals respectively inhibit two transcription factors, 
the second of which regulates the fluorescent output.  We consider all ways in which each of the transcription factors may regulate itself and the other (with the constraint that neither is unregulated) and distinguish between up- and down-regulation, giving a total of $160$ networks (the 
combinatorial accounting is presented in Sec.\ \ref{sec:comb}).  
The topology of each network $n \in \{1, 2, \dots 160\}$ constrains the model parameters $\th$ to lie within the particular feasibility set $\Omega_n$.  Having defined the set of networks, we enumerate topological features and their values $v_\mu$ (here, $\mu\in\{1,2,\dots 17\}$; see Table \ref{rhotable}).

\subsection{Regulatory model}

The mean protein numbers of the two transcription factors $\bar{A}$ and $\bar{B}$ and the fluorescent output $\bar{G}$ are described by the deterministic dynamics
\beqn
\frac{1}{R_A} \frac{d\bar{A}}{dt} &=& \varphi_A(a,b) - \bar{A}, \nonumber \\
\label{eq:dynsys}
\frac{1}{R_B} \frac{d\bar{B}}{dt} &=& \varphi_B(a,b) - \bar{B}, \\
\frac{1}{R_G} \frac{d\bar{G}}{dt} &=& \varphi_G(b) - \bar{G}, \nonumber
\eeqn
where $a = \{\bar{A}/x,\bar{A}\}$ when the first inhibitor is $\{$present, absent$\}$, $b = \{\bar{B}/y,\bar{B}\}$ when the second inhibitor is $\{$present, absent$\}$,
and the $R_j$ are degradation rates ($j \in \{A,B,G\}$).
The parameters $x > 1$ and $y > 1$ model
the effect of the interfering small molecules
by reducing the effective concentrations of the proteins.
This gives a total of four chemical input states 
denoted $c \in \{--,-+,+-,++\}$, each state describing whether
each of the two inhibitors is present $(+)$ or absent $(-)$.
The terms $\varphi_j$
describe the transcriptional regulation of each species by its parent(s) and are formulated under a statistical mechanical model \cite{Buchler, Bintu1, Bintu2}.  The statistical mechanical approach to modeling transcription is principled, compact, and in the case of combinatorial regulation \cite{Buchler} captures the diversity of multidimensional reponses observed in experimental systems \cite{Kaplan, Cox, Mayo}.  Full algebraic forms of the $\varphi_j$ are dependent on topology, including, in the case of combinatorial regulation, whether the transcription factor interaction is additive or multiplicative (see Sec.\ \ref{sec:reg}).

The stochastic description of each network is set by intrinsic noise.  We obtain probability distributions over protein numbers using the linear noise approximation (LNA) \cite{Elf,Paulsson,vanKampen}, since, in contrast to simulation techniques \cite{Gillespie}, the LNA does not rely on sampling and is thus much more
computationally efficient (making many-parameter optimization feasible).  Under the LNA the steady state distribution over each species' protein number is a Gaussian expansion around the deterministic mean given by the steady state of Eqn.\ \ref{eq:dynsys}.
The covariance matrix $\Xi$ under the LNA is determined from model parameters by (numerically) solving the Lyapunov equation $J\Xi+\Xi J^T+D=0$, where $J$ is the Jacobian of the system in Eqn.\ \ref{eq:dynsys} and
$D = {\rm diag}\{R_A(\varphi_A+\bar{A}), R_B(\varphi_B+\bar{B}),R_G(\varphi_G+\bar{G})\}$
is an effective diffusion matrix.
Of particular 
importance are the distributions $p(G|c)$, the stochastic response of the output species $G$ given that the system is in each of the four input states $c$.  The input-output MI may be computed directly from this quantity, $I[p(c,G)] = \sum_c \int dG\, p(G|c) p(c) \log [p(G|c)/\sum_{c'} p(G|c') p(c')]$, with the provision that the input states are equally likely, $p(c) = 1/4$.

\subsection{Input-output functions}

The possible input-output responses of each network are found by locally optimizing MI in parameter space.
The optimization is done numerically using MATLAB's fminsearch and initialized by sampling uniform-randomly in the logs of the parameters (specific bounds from which initial parameters are sampled are given in Table \ref{boundtable}).
The optimization is performed at constrained average protein number
$N \equiv (\bar{A}+\bar{B}+\bar{G})/3$
and average timescale separation $T \equiv [(R_A+R_B)/2]/R_G$ by maximizing the quantity $L \equiv I - \eta N - \kappa T$ for various values of the Lagrange multipliers $\eta$ and $\kappa$ ($R_G$ is fixed).

Optimization of MI has the effect of increasing the separation among the distributions $p(G|c)$ (see Fig.\ \ref{fig:cartoon}B-C).
To reflect the fact that many observed regulatory networks are known to 
operate near their information-optimal limits 
\cite{Tkacik2, Mehta, WalczakII}, we use in the statistical analysis only those optimal solutions whose MI lies above a cutoff value.
Choosing a cutoff larger than $1.5$ bits (which corresponds to two fully separated distributions and two fully overlapping distributions) ensures that the means of the distributions are fully resolved, and thus allows one to define the function performed by the network
as the ranking $r$ of the means of the distributions $p(G|c)$ along the $G$ axis.
For the results in this study we use a cutoff of $1.55$ bits.
The method can be easily extended to include less informative locally optimal functions, e.g.\ binary logic gates such as an AND function, by using a lower MI cutoff and generalizing the definition of function as ranking \cite{Mugler1}.
Fig.\ \ref{fig:cartoon}B-C
shows examples of two different functions performed by the same network that are local optima in MI at different points in parameter space; they correspond to $r=1$ and $r=9$ on the horizontal axis of Fig.\ \ref{fig:conditionals}, respectively.

To correct for repeated sampling of the same local optimum at close but numerically distinct points in the real-valued parameter space, nearest-neighbor optima performing the same function are merged.
This enforces that the distribution of optimal parameters is sampled uniformly.  The robustness of subsequent results to this choice is tested numerically (see Sec.\ \ref{sec:robust}).

\subsection{Correlating feature value and function}

For each topological feature $\mu$, the correlation between feature value $v_\mu$ and function $r$ is computed from the joint probability distribution $p(v_\mu,r)$.  This distribution is 
defined by the optimization data and the factorization
\beqn
p(v_\mu,r) &=& \sum_{\th,n} p(v_\mu,r,\th,n)
	= \sum_{\th,n} p(v_\mu,r|\th,n)p(\th,n) \nonumber \\
\label{eq:pvr}
	&=& \sum_{\th,n} p(v_\mu|n)p(r|\th,n)p(\th|n)p(n),
\eeqn
where $\th$ runs over all points in parameter space at which an optimum is found.
Here $p(v_\mu|n)$ is $\{0,1\}$-valued, set by whether network $n$ has value $v$ for feature $\mu$; and $p(r|n,\th)$ is $\{0,1\}$-valued, set by whether network $n$ performs function $r$ at point $\th$, according to the optimization data.
The distributions $p(\th|n)$ and $p(n)$ are assumed to be ``flat,'' i.e.\ $p(\th|n) = 1/|\th|_n$ and $p(n) = 1/|n|$, where $|\th|_n$ is the number of distinct local optima in parameter space
for network $n$, and $|n|$ is the number of networks; the robustness of subsequent results to weakening either of these assumptions is tested numerically (see Sec.\ \ref{sec:robust}).

The correlation between feature value and function is 
computed as their MI,
normalized by the entropy of $p(v_\mu)$, to yield a statistic
\beq
\rho_\mu \equiv \frac{I[p(v_\mu,r)]}{H[p(v_\mu)]}
\eeq
that ranges from $0$ (when the function provides no information about the feature value) to $1$ (when the feature value is exactly determined by the function).

\section{Results}

\subsection{Non-parametric analysis}\label{sec:nonpar}

\begin{figure}
\begin{center}
\includegraphics[scale=.47]{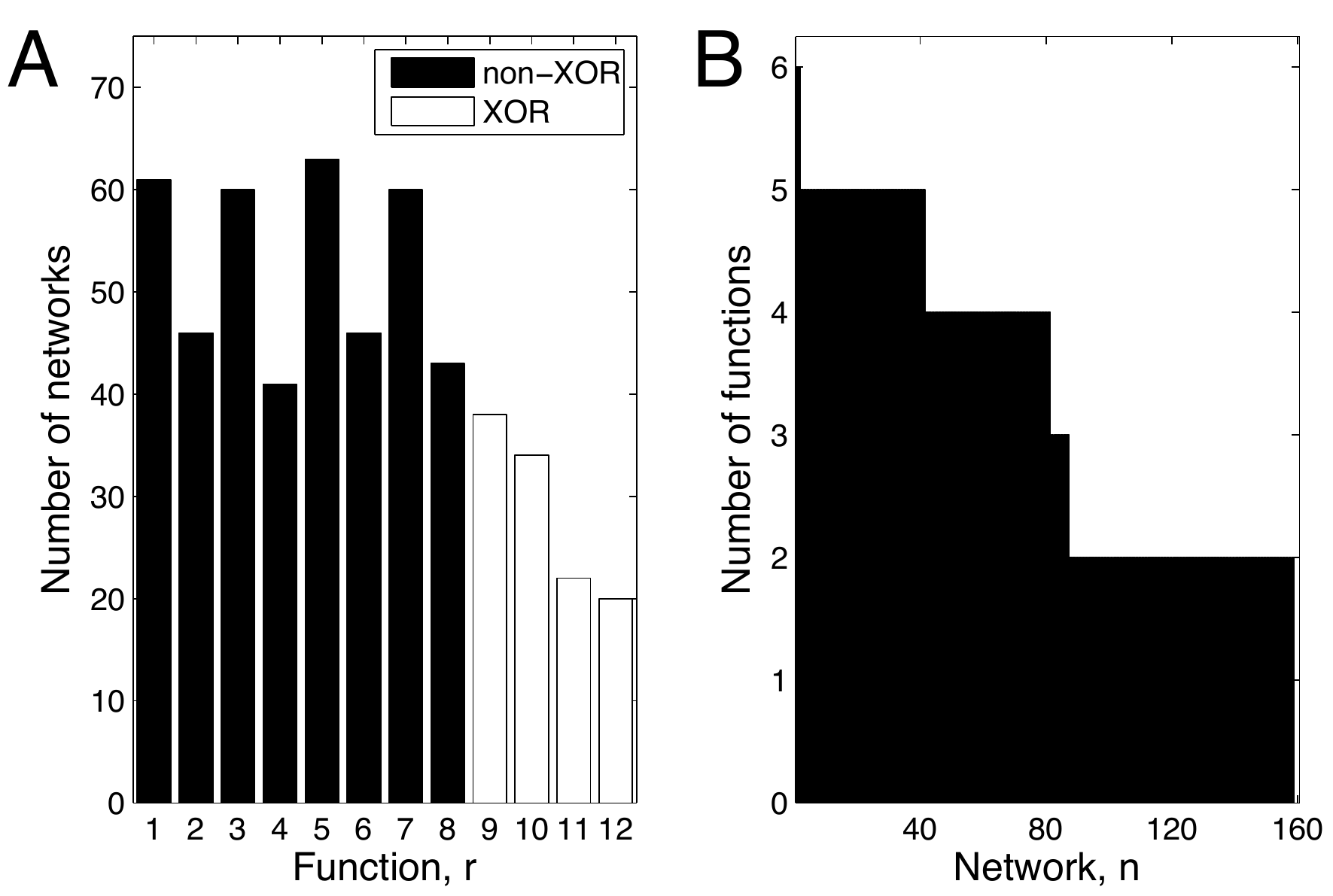}
\caption{Non-parametric analysis of network functionality.  (A) Histogram showing how many networks can perform each \IO function.  Functions are numbered along the horizontal axis as in Fig.\ \ref{fig:conditionals}.  (B) Histogram showing how many \IO functions may be realized by each network.  The order of networks along the horizontal axis is determined by ranking according to number of functions realized.}
\label{fig:histograms}
\end{center}
\end{figure}

We first present an analysis which requires no assumptions about 
what is a ``flat" distribution in parameter space, i.e., we 
simply enumerate how many networks are capable of a given 
degeneracy of \IO functions, and how many \IO functions may be 
realized by a given multiplicity of networks (Fig.\ \ref{fig:histograms}). This analysis recovers 
an intuitive result (Fig.\ \ref{fig:histograms}A): that ``XOR" functions, in which the sign 
of the influence of one input depends on the value of the other, are more
difficult to realize (i.e.\ they are observed in fewer networks). Simpler functions, 
in which the influence of one input remains positive or negative irrespective of the 
value of the other, are easier to realize.
The analysis also reveals that each network can perform at least two functions (Fig.\ \ref{fig:histograms}B).  These are the the two functions consistent with the signs of the forward regulatory edges $A \rightarrow B$ and $B \rightarrow G$, as described in detail in Sec.\ \ref{sec:forward}.  
Since the topology $A \rightarrow B \rightarrow G$
is that obtained in the parametric limit when the feedback edges are of negligible strength, it is clear that these functions must be realizable; Fig.\ \ref{fig:histograms}B shows further that these functions are sufficiently informative to be observed as information-optima.

\subsection{Topological features and robustness}\label{sec:robust}

Table \ref{rhotable} ranks the topological features by $\rho$, which measures how uniquely form determines function.  Recall that in computing $\rho$ from $p(v_\mu,r)$ we assume that the distributions $p(n)$ and $p(\th|n)$ are both uniform (Eqn.\ \ref{eq:pvr}); we find that the ranking in Table \ref{rhotable} is robust to deviations of both distributions from uniformity, as demonstrated by the following numerical experiments.

\begin{table}
\begin{tabular}{|l |l |}
\hline
Topological feature & $\rho$ \\
\hline
1. Sign of forward edges & $0.92366$ \\
2. Number of up-regulating edges & $0.18249$ \\
3. Number of down-regulating edges & $0.18098$ \\
4. Sign of autoregulation of species $B$ & $0.03228$ \\
5. Number of positive feedback cycles & $0.01133$ \\
6. Number of negative feedback cycles & $0.01109$ \\
7. Nesting structure of feedback cycles & $0.00418$ \\
8. Type of interaction of edges into $B$ & $0.00288$ \\
9. Number of edges & $0.00184$ \\
10. Number of feedback cycles & $0.00184$ \\
11. Sign of $A$-$B$ feedback cycle & $0.00173$ \\
12. Number of additive interactions & $0.00141$ \\
13. Type of interaction of edges into $A$ & $0.00085$ \\
14. Number of nested feedback cycles & $0.00081$ \\
15. Sign of autoregulation of species $A$ & $0.00058$ \\
16. Number of multiplicative interactions & $0.00048$ \\
17. Sign of edge $B \rightarrow A$ & $0.00021$ \\
\hline
\end{tabular}
\caption{Topological features ranked by correlation measure $\rho$.}
\label{rhotable}
\end{table}

The uniformity of $p(n)$ is perturbed by artificially setting $p(n) \propto (u_n)^\epsilon$, where $u_n$ is a vector of random numbers and $\epsilon$ tunes the entropy of the distribution, i.e.\ $\epsilon=0$ recovers the maximum-entropy (uniform) distribution, while $\epsilon \rightarrow \infty$ produces the zero-entropy distribution $p(n)=1 \Leftrightarrow n={\rm argmax~}(u_n)$.
We find that the ranking of the top $4$ features is preserved under $\sim$$15\%$ perturbations in the entropy, and that the ranking of the top $3$ features is preserved under $\sim$$30\%$ perturbations (see Fig.\ \ref{fig:robust}A).  This demonstrates that the feature ranking is considerably robust to perturbations in the uniformity of $p(n)$.

The uniformity of $p(\th|n)$ is perturbed similarly, and we find that the ranking of the top $7$ features is preserved under $\sim$$40\%$ perturbations in the entropy of $p(\th)$ (see Fig.\ \ref{fig:robust}B).  In this case we also have an independent entropy scale, given by the fact that we may decompose $p(\th|n)$ as $p(\th|n) = \sum_{\th_0} p(\th|\th_0,n)p(\th_0|n)$, where $\th_0$ is the parameter setting that initializes an optimization and $p(\th|\th_0,n)$ is determined by the optimization itself.  If we assume uniformity of $p(\th_0|n)$, instead of $p(\th|n)$, then $p(\th|n)$ is computable from the numbers of times the optimization converges repeatedly on each local optimum $\th$.  The entropy in this case is $13\%$ different from that of the uniform distribution, and the ranking of $\rho$ is almost entirely unchanged (Fig.\ \ref{fig:robust}B).
This observation demonstrates that the results are not sensitive to whether one takes the distribution of initial parameters or the distribution of optimal parameters to be uniform.

\subsection{Non-redundant features}

Many topological features are not independent; 
for example, the feature ``number of up-regulating edges'' is
highly correlated with ``number of down-regulating edges.''
To interpret which features are associated with which sets of
realizable functions, it is useful to group nearly identical features together
and use only the feature which is most informative about function
(highest in $\rho$) as the exemplar among each group.
To quantify redundancy among features, we compute the MI between each pair of features and normalize by the minimum entropy to produce a weighted adjacency matrix $M_{\mu\nu} = I[p(v_\mu,v_\nu)]/\min\{H[p(v_\mu)],H[p(v_\nu)]\}$, which
we then use as the basis for unidimensional scaling \cite{borg} (see Sec.\ \ref{sec:MDS}).

\begin{figure}
\begin{center}
\includegraphics[scale=.47]{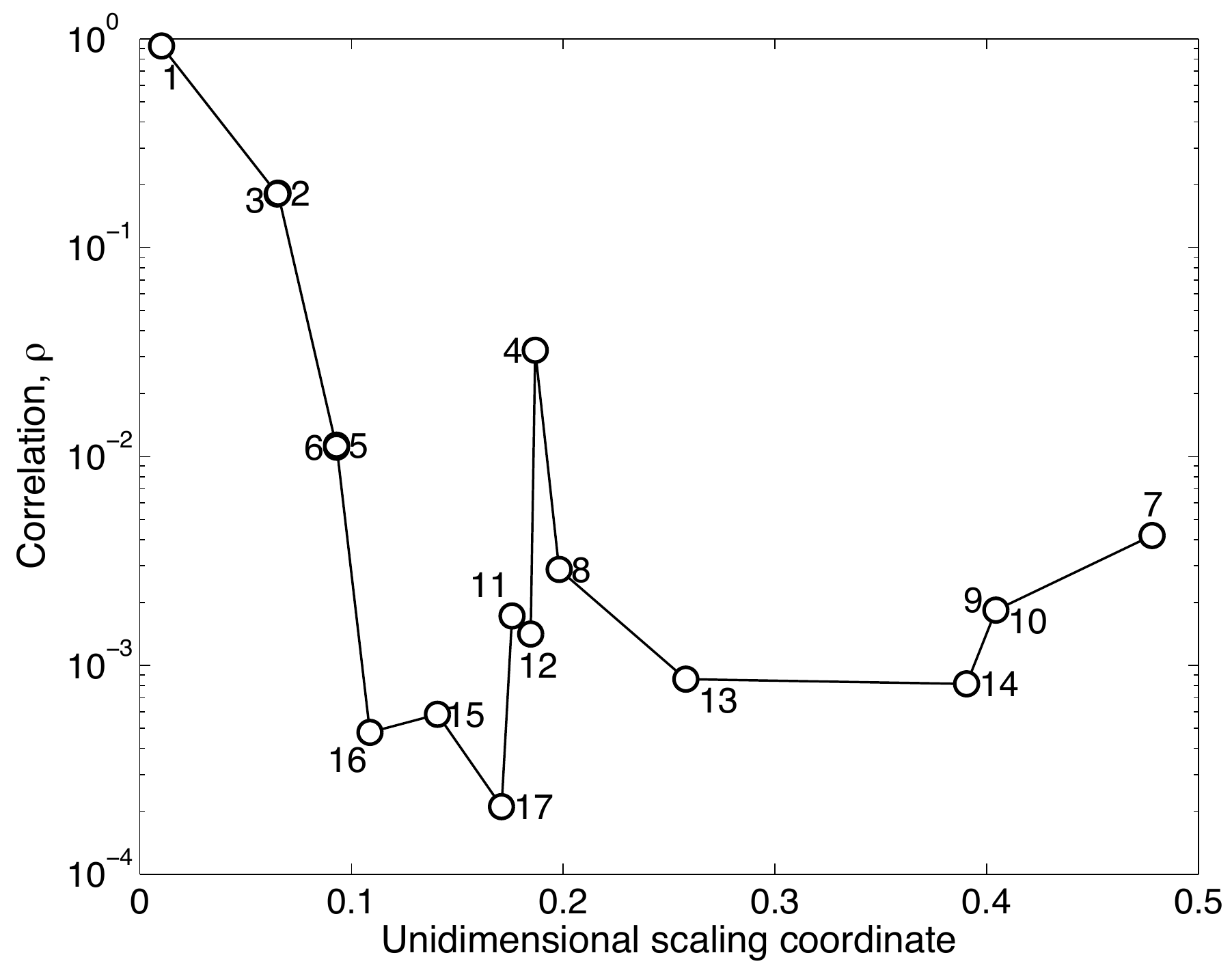}
\caption{Identifying feature redundancy.  Correlation measure $\rho$ is plotted against a unidimensional scaling coordinate which groups similar features together (i.e.\ the components of the eigenvector corresponding to the largest-magnitude eigenvalue of the feature adjacency matrix $M_{\mu\nu}$).  Features are numbered by rank (Table \ref{rhotable}).}
\label{fig:redundancy}
\end{center}
\end{figure}

\begin{figure}
\begin{center}
\includegraphics[scale=0.37]{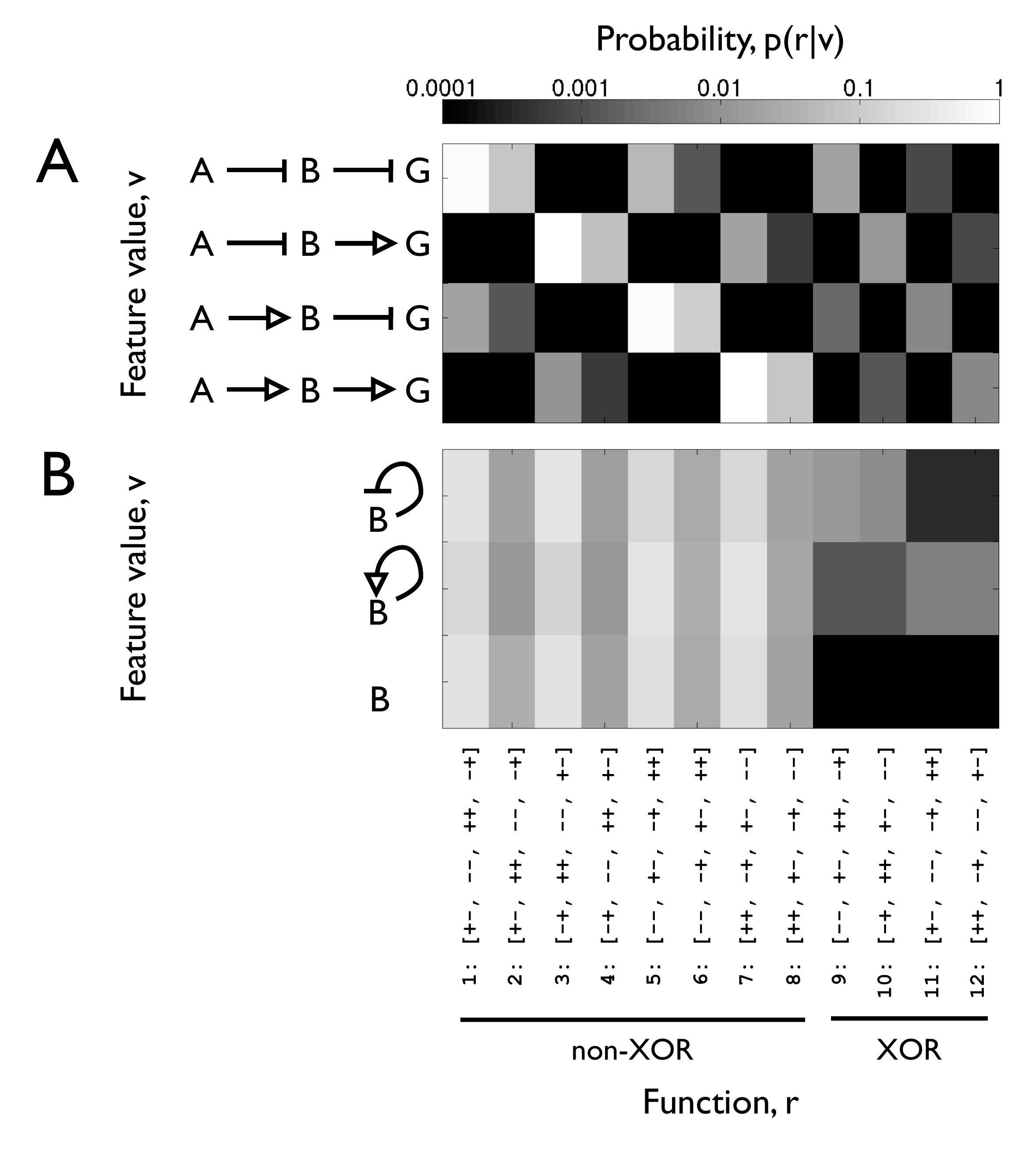}
\caption{Conditional distributions showing the probability of a particular input-output function given the value of a topological feature, for two features: (A) the signs (up- or down-regulating) of the forward regulations, and (B) the sign (up-regulating, down-regulating, or absent) of the autoregulation of species $B$.}
\label{fig:conditionals}
\end{center}
\end{figure}

Fig.\ \ref{fig:redundancy} plots features' $\rho$ values against the unidimensional scaling coordinate, revealing two distinct groups of highly informative features.  The first, which includes the features ranked $1$, $2$, $3$, $5$, and $6$, is dominated by feature $1$: the signs (up- or down-regulating) of the forward regulatory edges $A \rightarrow B$ and $B \rightarrow G$.  The second, which includes the features ranked $4$, $8$, $11$, and $12$, is dominated by feature $4$: the sign (up-regulating, down-regulating, or absent) of the autoregulation of species $B$.  The high information content of each of these two features is revealed visually by inspection of the conditional distribution $p(r|v_\mu) = p(v_\mu,r)/p(v_\mu)$ (Fig.\ \ref{fig:conditionals}), as described in detail in the next sections.  The functional importance of both of these features is supported by analytic results; for the first feature these analytic predictions were made in earlier work \cite{Mugler1} and are recalled here, while for the second feature new analytic results are derived here.

\subsection{Forward regulation}\label{sec:forward}

The topological feature that is most informative of network function is feature 1: the signs of the forward regulatory edges $A \rightarrow B$ and $B \rightarrow G$.  Inspection of the feature value--function conditional distribution $p(r|v)$ in Fig.\ \ref{fig:conditionals}A reveals a rich, highly organized (and thus highly informative) structure which we here interpret.

The most prominent aspect of the probability matrix in Fig.\ \ref{fig:conditionals}A is the high-probability double-diagonal spanning the eight non-XOR functions (i.e. functions $1$ and $2$ are most often performed by networks with the first feature value, functions $3$ and $4$ the second feature value, functions $5$ and $6$ the third feature value, and functions $7$ and $8$ the fourth feature value).  These are the functions one would expect by looking at the forward edges alone, i.e.\ in the absence of feedback.  For example, in networks with the last feature value, $A \rightarrowtriangle B \rightarrowtriangle G$, inhibition of $A$ and of $B$ will both reduce the expression of $G$, such that the state in which both small molecules are present ($++$) produces the lowest-ranked output, and conversely, the state in which both small molecules are absent ($--$) produces the highest-ranked output; functions $7$ and $8$ are the two that satisfy these criteria.  In previous work \cite{Mugler1} we termed these functions ``direct,'' and we showed analytically that networks are limited to direct functions even when feedback is added, so long as each species is regulated by at most one other species.  This fact is validated here numerically: a plot of $p(v|r)$ for only those networks in our set in which each species is regulated by one other species shows nonzero entries only for the direct functions (Fig.\ \ref{fig:valid}).

Among all networks, including those with combinatorial feedback (i.e.\ two edges impinging on one node), we see that direct functions still dominate, indicated by the bright double-diagonal in Fig.\ \ref{fig:conditionals}A.  Networks with combinatorial feedback perform other functions as well, but more rarely; examples include those functions in Fig.\ \ref{fig:conditionals}A above and below the double-diagonal and XOR functions $9$-$12$.  The performance of these additional functions remains well organized by feature value, which makes the signs of the forward edges a highly informative feature.

\subsection{Autoregulation of $B$}

Other than feature 1 (and the features highly correlated with feature 1), the most informative feature is feature 4: the autoregulation of species $B$.
Inspection of its feature value--function distribution (Fig.\ \ref{fig:conditionals}B) reveals that the information content lies in the ability to perform XOR functions.  Specifically, networks in which $B$ is autoregulated are observed to perform XOR functions, while networks in which $B$ is not autoregulated are not observed to perform XOR functions.  Indeed, autoregulation has been observed to enhance the functional response to multiple inputs in a related study in the context of Boolean logic gates \cite{Hermsen}.

XOR functions are those in which the sign of the influence of one input depends on the value of the other; mathematically they satisfy one or both of two properties:
\beqn
{\rm XOR \,\, property \,\, I:} &{\rm sign} \left( d\bar{G}/dx \right) {\rm \,\, depends \,\, on \,\,} y,& \nonumber \\
{\rm XOR \,\, property \,\, II:} &{\rm sign} \left( d\bar{G}/dy \right) {\rm \,\, depends \,\, on \,\,} x,& \nonumber
\eeqn
The four observed XOR functions satisfy property I (e.g.\ Fig.\ \ref{fig:cartoon}C inset); no functions satisfying property II are observed (Fig.\ \ref{fig:conditionals}B).  Analytic support for these facts is obtained by calculating $d\bar{G}/dx$ and $d\bar{G}/dy$, respectively.

To understand why autoregulation of $B$ is necessary for XOR functions satisfying property I, we calculate $d\bar{G}/dx$ analytically.  We obtain (see Sec.\ \ref{sec:deriv})
\beq
\label{eq:dGdx}
\frac{d\bar{G}}{dx} = \frac{1}{-\Delta}\frac{\partial a}{\partial x}
	\frac{\partial \varphi_B}{\partial a}\frac{\partial \varphi_G}{\partial \bar{B}},
\eeq
where
$\Delta=(\partial \varphi_A / \partial \bar{B})(\partial \varphi_B / \partial \bar{A}) - [(\partial \varphi_A / \partial \bar{A}) - 1][(\partial \varphi_B / \partial \bar{B}) - 1]$
is the determinant of the Jacobian of the dynamical system in Eqn.\ \ref{eq:dynsys} and is always negative for stable fixed points.  Eqn.\ \ref{eq:dGdx} has an intuitive form when considering the direct path from $x$ to $G$ (Fig.\ \ref{fig:cartoon}A): the term $\partial a / \partial x = -\bar{A}/x^2 < 0$ corresponds to the inhibitory signal $x \leadsto A$, the term $\partial \varphi_B / \partial a$ corresponds to the regulatory edge $A \rightarrow B$, and the term $\partial \varphi_G / \partial \bar{B}$ corresponds to the regulatory edge $B \rightarrow G$.  Since $G$ has only one regulatory input, $\varphi_G(b)$ is monotonic, making $\partial \varphi_G / \partial \bar{B} = (d \varphi_G / d b)(\partial b / \partial \bar{B}) = (d \varphi_G / d b)/y$ of unique sign.  The same is true for $\partial \varphi_B / \partial a$ when $B$ has only one regulatory input (i.e.\ when $B$ is not autoregulated). However when $B$ has more than one regulatory input (i.e. when $B$ is autoregulated), the sign of $\partial \varphi_B / \partial a$ can depend on $y$, allowing XOR property I.  Specifically, under our regulatory model, when $B$ is autoregulated, $\partial \varphi_B / \partial a$ is the product of a positive term and a term quadratic in $b = \bar{B}/y$ that has positive roots for some parameter settings (see Sec.\ \ref{sec:prop1}).
This analysis suggests inspection of the parameters themselves obtained via optimization; doing so, we observe that the vast majority of observed XOR functions result from optimal parameter values for which there exists a positive root in the range $0 < \bar{B}/y < $ $\sim$$100$, which is the range of protein numbers in which our optimal solutions lie
(Fig.\ \ref{fig:cartoon}B-C).
To summarize, nonmonotonicity in the regulation of species $B$, which can occur only when $B$ is autoregulated, produces the observed XOR functions.

To understand why XOR functions satisfying property II are not observed, we calculate $d\bar{G}/dy$ analytically.  We obtain (see Sec.\ \ref{sec:deriv})
\beq
\label{eq:dGdy}
\frac{d\bar{G}}{dy} = \frac{1}{-\Delta} \left( 1 - \frac{\partial \varphi_A}{\partial \bar{A}} \right)
	\frac{\partial b}{\partial y}\frac{d \varphi_G}{d b},
\eeq
where as before the determinant $\Delta$ is always negative.  The last two terms correspond to edges along the direct path from $y$ to $G$, i.e.\ $y \leadsto B$ and $B \rightarrow G$ respectively, and are of unique sign; the term in parentheses describes the effect of the upstream species $A$ and feedback.
In all optimal solutions the term in parentheses is observed to be positive, despite wide variations in the orders of magnitude of each of the optimal parameters across solutions.  This observation is largely explained by a stability analysis: for four of the six possible topological classes of networks (those in which $A$ is singly, not doubly, regulated; Fig.\ \ref{fig:cartoon}A), stability of a fixed point of Eqn.\ \ref{eq:dynsys} implies that the term in parentheses is positive; for the other two topological classes, stability implies that this term is greater than a quantity that is zero for some parameter settings and of unknown sign for others (see Sec.\ \ref{sec:prop2}).  In this last case it is unclear whether negative values of this term are analytically forbidden or simply exceedingly unlikely given the regulatory model and the space of optimal solutions.  Empirically this term is always positive, and type-II XOR functions are not observed.

The necessity of both forward regulation and autoregulation of species $B$ for the performance of XOR functions highlights the importance of combinatorial regulation in functional versatility.
As previously mentioned, networks without combinatorial regulation are limited to a particular class of functions which does not include XOR functions \cite{Mugler1}.
Moreover, in the original experiment of Guet et al.\ \cite{Guet}, each species was singly regulated, and accordingly no XOR functions were observed.

\section{Discussion}

Both in order to assign functional significance to observed small network topologies in nature,
and to design synthetic networks which will execute a desired function or set of functions, it is useful
to develop a systematic approach for revealing the extent to which the form of a small network
guides or constrains its functions. Resorting to hypothesized functions may be appealing
in terms of interpretability, but this strategy risks overemphasizing those functions which
one is looking for, or overlooking an unexpected function entirely.

The statistical analysis, along with the analytic results presented above, illustrate how the
search for form-function relations can be posed as an algorithmic approach leading to interpretable
mechanisms. While we have illustrated it for a particular, experimentally-realized setup, the
approach itself, subdivided into a set of distinct modules in Fig. \ref{fig:pipeline}, we anticipate
will be applicable to a wide variety of biophysical contexts. Similarly, we have chosen a
framework from which much can be discovered by analytic study of the deterministic dynamical system;
other experimental setups may require vastly different analytic explanations, but the idea
of using statistical methods to highlight the features of paramount importance should be
implementable as illustrated. We look forward to exploring the extent to which form does
or does not follow function --- and how --- in related biophysical and biochemical models of small 
information processing networks in biology.

\section{Appendix}\label{sec:app}

The remainder of this paper contains details on the set of networks studied, the regulatory model, the optimization procedure, and the statistical analysis used to identify exemplars among non-redundant groups of topological features.  Additionally, it includes further background on information theory and the linear noise approximation.  Finally, it presents numerical and analytic results as referenced in the main text, including numerical tests of the robustness of results to uniformity assumptions made in the statistical analysis, statistical validation of an analytically derived functional constraint, and analytic results on the ability of networks to perform XOR functions.

\subsection{Networks}\label{sec:comb}

Many researchers over the past decades have begun to focus on the description of a given biological system not in terms of the isolated functions of its independent components, but in terms of the collective function of the network of interacting components as a whole.  A central example of such a network is that describing interactions among genes: many genes produce proteins called transcription factors, whose role is to influence the protein production of other genes.  The goal of the present study is, for a set of such networks, to develop a statistical method for determining the extent to which the topology of the network correlates with the function(s) it performs.  In this section we describe the set of networks considered, including presentation of its combinatorics.

We consider the set of networks consisting of two transcription factors $A$ and $B$, each of which can be chemically inhibited, each of which is regulated by itself, the other, or both, and one of which regulates the expression of a fluorescent output gene $G$ (Fig.\ 1A).  We distinguish between up- and down-regulating edges, and, in the case of more than one regulatory input edge, between additive and multiplicative interaction of the transcription factors (see model in next section).  This gives a set of $160$ networks, as described by the combinatorics here.

\begin{figure}
\begin{center}
\includegraphics[scale=.25]{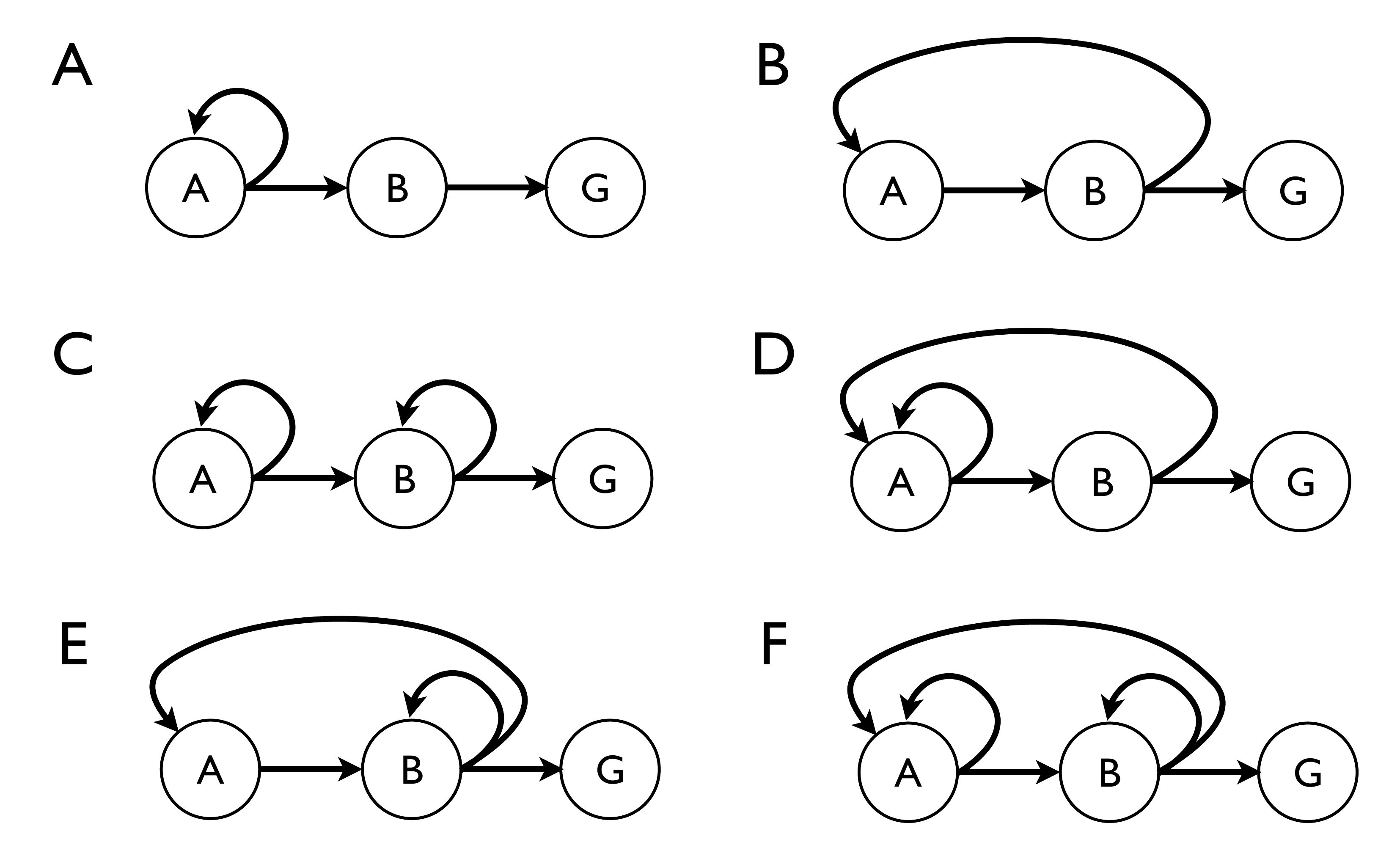}
\caption{The six possible topological configurations of the networks studied.}
\label{fig:topos}
\end{center}
\end{figure}

There are six ways in which the three possible feedback edges illustrated in Fig.\ 1A can appear such that species $A$ remains regulated.  In $2$ configurations (Figs.\ \ref{fig:topos}A-B), both $A$ and $B$ are singly regulated, and there are a total of $3$ edges per configuration, each of which can be up- or down-regulating.  The number of networks is thus [number of configurations]$\times$[$2$\verb|^|(number of edges)] $= [2]\times [2^{(3)}] = 16$.

In $3$ configurations, (Fig.\ \ref{fig:topos}C-E), one of $A$ and $B$ is singly regulated while the other is doubly regulated.  In the case of double regulation, the interaction between transcription factors is either (i) additive, in which the signs (up or down) of the two regulatory edges are independent, giving $4$ possibilities, or (ii) multiplicative, in which the two regulatory edges have the same sign, giving $2$ possibilities (see model in next section), for a total of $6$ possibilities.  The number of networks is thus [number of configurations]$\times$[$2$\verb|^|(number of singly regulated nodes)]$\times$[$6$\verb|^|(number of doubly regulated nodes)] $= [3]\times [2^{(2)}]\times [6^{(1)}] = 72$.

In the last configuration (Fig.\ \ref{fig:topos}F), both $A$ and $B$ are doubly regulated, and the number of networks is [number of configurations]$\times$[$2$\verb|^|(number of singly regulated nodes)]$\times$[$6$\verb|^|(number of doubly regulated nodes)] $= [1]\times [2^{(1)}]\times [6^{(2)}] = 72$.  This gives a total of $16+72+72=160$ networks.

\subsection{Gene regulation}

A regulatory network is most simply described using a dynamical system whose degrees of freedom are the concentrations of its constituent proteins.  The nature of the interactions is then determined by the functional form of regulatory production terms on the righthand side of the dynamical system.  In this study we formulate production terms under a statistical mechanical model \cite{Buchler, Bintu1, Bintu2} that has been shown to capture diverse functionality in the case of combinatorial regulation \cite{Buchler}.

Because many proteins are present in cells in as few as tens or hundreds of copies \cite{Guptasarma}, a deterministic dynamical system, which ignores intrinsic fluctuations around mean protein concentrations, can provide an insufficient description of the biological system when copy numbers are low.  Ideally one instead seeks to solve the {\it master equation}, which describes the dynamics of the {\it probability} of observing given protein numbers.  Unfortunately, almost all master equations describing regulatory networks, including the set of small networks we study here, are intractable analytically.  As a result, several techniques to simulate \cite{Gillespie} or approximate \cite{vanKampen} the master equation have been developed; we here employ the {\it linear noise approximation} (LNA).  Since the LNA does not rely on sampling (in contrast to simulation techniques), it is computationally efficient, which makes feasible the many-parameter optimization performed in this study.

The LNA is a second-order expansion of the master equation, with the first-order terms recovering the deterministic description.  Therefore, in the present section, we first describe in detail the regulatory model used in the deterministic system; then we describe the LNA and its application to our networks.

\subsubsection{Deterministic description: Dynamical system}\label{sec:reg}

As in Eqn.\ $1$, the mean protein numbers of the three species are described by the following deterministic dynamics (for notational brevity the bars have been dropped --- i.e.\ the quantities $\bar{A}$, $\bar{B}$, and $\bar{G}$ have been changed to $A$, $B$, and $G$ --- and the regulation terms $\varphi_A$, $\varphi_B$, and $\varphi_G$ have been relabeled $\alpha$, $\beta$, and $\gamma$, respectively):
\beqn
\label{eq:dyn1}
\frac{1}{R_A} \frac{dA}{dt} &=& \alpha(a,b) - A, \\
\label{eq:dyn2}
\frac{1}{R_B} \frac{dB}{dt} &=& \beta(a,b) - B, \\
\label{eq:dyn3}
\frac{1}{R_G} \frac{dG}{dt} &=& \gamma(b) - G,
\eeqn
where $a = \{A/x,A\}$ when the first inhibitor is $\{$present, absent$\}$, $b = \{B/y,B\}$ when the second inhibitor is $\{$present, absent$\}$,
and the $R_j$ are degradation rates ($j \in \{A,B,G\}$).
The parameters $x > 1$ and $y > 1$ incorporate the inhibition
by reducing the effective concentrations of the proteins.
This gives a total of four chemical input states $c \in \{--,-+,+-,++\}$, each state describing whether each of the two inhibitors is present $(+)$ or absent $(-)$.

\paragraph{Stability}

Steady state solutions of the dynamical system in Eqns.\ \ref{eq:dyn1}-\ref{eq:dyn3} are stable fixed points, i.e.\ points at which the all eigenvalues of the Jacobian have negative real part.  The Jacobian of the system is
\beq
\label{eq:Jac}
J = \begin{pmatrix}
\pd{\alpha}{A}-1 & \pd{\alpha}{B} & 0 \\
\pd{\beta}{A} & \pd{\beta}{B}-1 & 0 \\
0 & \pd{\gamma}{B} & -1 \\
\end{pmatrix}.
\eeq
Its eigenvalues are
\beq
\label{eq:eig}
\lambda_{1,2} = \frac{1}{2} \left[ \pd{\alpha}{A} + \pd{\beta}{B} - 2 \pm
	\sqrt{ \left( \pd{\alpha}{A} - \pd{\beta}{B} \right)^2 + 4 \pd{\alpha}{B} \pd{\beta}{A} } \right]
\eeq
and $\lambda_3 = -1$.  At stable fixed points ${\rm Real}\{\lambda_{1,2}\} < 0$; equivalently the determinant of Eqn. \ref{eq:Jac}, which is real and equal to the product of the eigenvalues, must be negative:
\beqn
\label{eq:det}
\Delta &\equiv& \det(J) = \lambda_1 \lambda_2 \lambda_3 \nonumber\\
&=& \pd{\alpha}{B} \pd{\beta}{A}
	- \left( \pd{\alpha}{A} - 1 \right) \left( \pd{\beta}{B} - 1 \right) < 0.
\eeqn

\paragraph{Regulation terms}

The regulatory model is that of Buchler et al.\ \cite{Buchler}, in which the protein production is proportional to the probability that the RNA polymerase (RNAp) is bound to the promoter of interest.  This probability is formulated thermodynamically, i.e.\ by enumerating and statistically weighting all ways in which transcription can and cannot occur.

For our networks, the regulation terms $\alpha$, $\beta$, and $\gamma$ are
\beqn
\label{eq:alpha}
\alpha(a,b) &=& \frac{s_A}{R_A} p_A(a,b), \\
\label{eq:beta}
\beta(a,b) &=& \frac{s_B}{R_B} p_B(a,b), \\
\label{eq:gamma}
\gamma(b) &=& \frac{s_G}{R_G} p_G(b),
\eeqn
where the $s_j$ are promoter strengths, and the probabilities of transcription $p_j$ are given by
\beq
\label{eq:prob}
p_j = \frac{Z_{\rm on}^j}{Z_{\rm on}^j+Z_{\rm off}^j}.
\eeq
The partition functions $Z_{\rm on}^j$ and $Z_{\rm off}^j$ describe all the ways that transcription can occur and not occur, respectively, at the promoter region of gene $j$.  As presented in detail below, the partition functions are determined by network topology and depend on additional parameters, including interaction strengths $w$, binding constants $K$, and ``leakiness''  $q$ (i.e.\ the statistical weight given to bare binding of the RNAp).  All parameters are positive.  We first offer qualitative interpretation of the parameters, then we present the detailed expressions for the partition functions.

The $w$ describe the interaction strengths between transcription factors, or between a transcription factor and the RNAp.  Alphabetical superscripts refer to the promoter regions of genes $A$, $B$, or $G$, while numerical subscripts refer to the molecules involved in the interaction: RNAp (0), transcription factor $A$ (1), or transcription factor $B$ (2).  For example, $w^B_{01}$ describes the interaction between RNAp and transcription factor $A$ at the promoter region of gene $B$.

The signs of the logs of the $w^j_{0i}$ determine the signs of the edges (where $j\in \{A,B,G\}$ for the three promoter regions and $i \in \{1,2\}$ for transcription factors $A$ and $B$).  For example, $w^B_{01}$ describes the regulation of species $B$ by species $A$; if $\log w^B_{01} > 0$ then the edge $A \rightarrow B$ is up-regulating, and if $\log w^B_{01} < 0$ then the edge $A \rightarrow B$ is down-regulating.

Following the model of Buchler et al. \cite{Buchler}, when two transcription factors regulate one species, they may do so ``independently'' or ``synergistically.''  Independent regulation corresponds to the case when both transcription factors interact with the same domain of the RNAp; mathematically the interaction strengths are additive ($w^j_{01}+w^j_{02}$).  If the RNAp has several interaction domains, two transcription factors can interact synergistically, and the interaction strengths are multiplicative ($w^j_{01}w^j_{02}$).  Synergistic regulation implies the additional constraint that the regulatory effects of the two transcription factors (i.e.\ the logs of the interaction strengths) are of the same sign.

The $K$ describe the binding constants of each transcription factor to each promoter region.  They have super- and subscripts similar to the $w$, e.g.\ $K^B_1$ describes the binding of transcription factor $A$ to the promoter region of gene $B$.

The partition functions $Z_{\rm on/off}^j$ for the network topologies shown in Fig.\ \ref{fig:topos} are presented below.
To provide intuition, the first two expressions are interpreted in detail.

Topology A is shown in Fig.\ \ref{fig:topos}A; its partition functions are
\beqn
\label{eq:ZAon}
Z_{\rm on}^A &=& q + w^A_{01} q \frac{a}{K_1^A}, \\
\label{eq:ZAoff}
Z_{\rm off}^A &=& 1 + \frac{a}{K_1^A}, \\
\label{eq:ZBon}
Z_{\rm on}^B &=& q + w^B_{01} q \frac{a}{K_1^B}, \\
\label{eq:ZBoff}
Z_{\rm off}^B &=& 1 + \frac{a}{K_1^B}, \\
\label{eq:ZGon}
Z_{\rm on}^G &=& q + w^G_{02} q \frac{b}{K_2^G}, \\
\label{eq:ZGoff}
Z_{\rm off}^G &=& 1 + \frac{b}{K_2^G}.
\eeqn
Eqn.\ \ref{eq:ZAon} describes the two ways in which transcription can occur at the promoter region of gene $A$: (i) the RNAp can bind unassisted, with statistical weight $q$, or (ii) since $A$ is self-regulating, the RNAp can bind upon interaction with transcription factor $A$, with weight proportional to $q$ for the RNAp, to the effective concentration $a$ scaled by the binding constant $K_1^A$ for transcription factor $A$, and to the interaction strength $w^A_{01}$ between the RNAp and transcription factor $A$.  Eqn.\ \ref{eq:ZAoff} describes the two ways in which transcription can not occur at the promoter region of gene $A$: (i) there can be nothing bound, an outcome whose statistical weight we are free, via the normalization enforced in Eqn.\ \ref{eq:prob}, to set, and so we set to $1$, and (ii) the transcription factor alone can bind, with weight $a/K_1^A$.  Eqns.\ \ref{eq:ZBon}-\ref{eq:ZGoff} are similarly derived according to the topology of the network.

Topology B is shown in Fig.\ \ref{fig:topos}B; its partition functions are
\beqn
Z_{\rm on}^A &=& q + w^A_{02} q \frac{b}{K_2^A}, \\
Z_{\rm off}^A &=& 1 + \frac{b}{K_2^A}, \\
Z_{\rm on}^B &=& q + w^B_{01} q \frac{a}{K_1^B}, \\
Z_{\rm off}^B &=& 1 + \frac{a}{K_1^B}, \\
Z_{\rm on}^G &=& q + w^G_{02} q \frac{b}{K_2^G}, \\
Z_{\rm off}^G &=& 1 + \frac{b}{K_2^G}.
\eeqn

Topology C is shown in Fig.\ \ref{fig:topos}C; its partition functions are
\beqn
Z_{\rm on}^A &=& q + w^A_{01} q \frac{a}{K_1^A}, \\
Z_{\rm off}^A &=& 1 + \frac{a}{K_1^A}, \\
\label{eq:ZBadd}
Z_{\rm on}^B &=& q + w^B_{01} q \frac{a}{K_1^B} + w^B_{02} q \frac{b}{K_2^B} \nonumber\\
	&& + (w^B_{01}+w^B_{02}) w^B_{12} q \frac{a}{K_1^B} \frac{b}{K_2^B} \quad {\rm (additive)}, \\
\label{eq:ZBmult}
Z_{\rm on}^B &=& q + w^B_{01} q \frac{a}{K_1^B} + w^B_{02} q \frac{b}{K_2^B} \nonumber\\
	&& w^B_{01} w^B_{02} w^B_{12} q \frac{a}{K_1^B} \frac{b}{K_2^B} \quad {\rm (multiplicative)}, \\
\label{eq:ZBoff2}
Z_{\rm off}^B &=& 1 + \frac{a}{K_1^B} + \frac{b}{K_2^B} + w^B_{12} \frac{a}{K_1^B} \frac{b}{K_2^B}, \\
Z_{\rm on}^G &=& q + w^G_{02} q \frac{b}{K_2^G}, \\
Z_{\rm off}^G &=& 1 + \frac{b}{K_2^G}.
\eeqn

Topology D is shown in Fig.\ \ref{fig:topos}D; its partition functions are
\beqn
Z_{\rm on}^A &=& q + w^A_{01} q \frac{a}{K_1^A} + w^A_{02} q \frac{b}{K_2^A} \nonumber\\
	&& + (w^A_{01} + w^A_{02}) w^A_{12} q \frac{a}{K_1^A} \frac{b}{K_2^A} \quad {\rm (additive)}, \\
Z_{\rm on}^A &=& q + w^A_{01} q \frac{a}{K_1^A} + w^A_{02} q \frac{b}{K_2^A} \nonumber\\
	&& + w^A_{01} w^A_{02} w^A_{12} q \frac{a}{K_1^A} \frac{b}{K_2^A} \quad {\rm (multiplicative)}, \\
Z_{\rm off}^A &=& 1 + \frac{a}{K_1^A} + \frac{b}{K_2^A} + w^A_{12} \frac{a}{K_1^A} \frac{b}{K_2^A}, \\
Z_{\rm on}^B &=& q + w^B_{01} q \frac{a}{K_1^B}, \\
Z_{\rm off}^B &=& 1 + \frac{a}{K_1^B}, \\
Z_{\rm on}^G &=& q + w^G_{02} q \frac{b}{K_2^G}, \\
Z_{\rm off}^G &=& 1 + \frac{b}{K_2^G}.
\eeqn

Topology E is shown in Fig.\ \ref{fig:topos}E; its partition functions are
\beqn
Z_{\rm on}^A &=& q + w^A_{02} q \frac{b}{K_2^A}, \\
Z_{\rm off}^A &=& 1 + \frac{b}{K_2^A}, \\
Z_{\rm on}^B &=& q + w^B_{01} q \frac{a}{K_1^B} + w^B_{02} q \frac{b}{K_2^B} \nonumber\\
	&& + (w^B_{01}+w^B_{02}) w^B_{12} q \frac{a}{K_1^B} \frac{b}{K_2^B} \quad {\rm (additive)}, \\
Z_{\rm on}^B &=& q + w^B_{01} q \frac{a}{K_1^B} + w^B_{02} q \frac{b}{K_2^B} \nonumber\\
	&& + w^B_{01} w^B_{02} w^B_{12} q \frac{a}{K_1^B} \frac{b}{K_2^B} \quad {\rm (multiplicative)}, \\
Z_{\rm off}^B &=& 1 + \frac{a}{K_1^B} + \frac{b}{K_2^B} + w^B_{12} \frac{a}{K_1^B} \frac{b}{K_2^B}, \\
Z_{\rm on}^G &=& q + w^G_{02} q \frac{b}{K_2^G}, \\
Z_{\rm off}^G &=& 1 + \frac{b}{K_2^G}.
\eeqn

Topology F is shown in Fig.\ \ref{fig:topos}F; its partition functions are
\beqn
Z_{\rm on}^A &=& q + w^A_{01} q \frac{a}{K_1^A} + w^A_{02} q \frac{b}{K_2^A} \nonumber\\
	&& + (w^A_{01} + w^A_{02}) w^A_{12} q \frac{a}{K_1^A} \frac{b}{K_2^A} \quad {\rm (additive)}, \\
Z_{\rm on}^A &=& q + w^A_{01} q \frac{a}{K_1^A} + w^A_{02} q \frac{b}{K_2^A} \nonumber\\
	&& + w^A_{01} w^A_{02} w^A_{12} q \frac{a}{K_1^A} \frac{b}{K_2^A} \quad {\rm (multiplicative)}, \\
Z_{\rm off}^A &=& 1 + \frac{a}{K_1^A} + \frac{b}{K_2^A} + w^A_{12} \frac{a}{K_1^A} \frac{b}{K_2^A}, \\
Z_{\rm on}^B &=& q + w^B_{01} q \frac{a}{K_1^B} + w^B_{02} q \frac{b}{K_2^B} \nonumber\\
	&& + (w^B_{01}+w^B_{02}) w^B_{12} q \frac{a}{K_1^B} \frac{b}{K_2^B} \quad {\rm (additive)}, \\
Z_{\rm on}^B &=& q + w^B_{01} q \frac{a}{K_1^B} + w^B_{02} q \frac{b}{K_2^B} \nonumber\\
	&& + w^B_{01} w^B_{02} w^B_{12} q \frac{a}{K_1^B} \frac{b}{K_2^B} \quad {\rm (multiplicative)}, \\
Z_{\rm off}^B &=& 1 + \frac{a}{K_1^B} + \frac{b}{K_2^B} + w^B_{12} \frac{a}{K_1^B} \frac{b}{K_2^B}, \\
Z_{\rm on}^G &=& q + w^G_{02} q \frac{b}{K_2^G}, \\
Z_{\rm off}^G &=& 1 + \frac{b}{K_2^G}.
\eeqn

\subsubsection{Stochastic description: Linear noise approximation}

The linear noise approximation (LNA) is a second-order expansion of the master equation made under the approximations that (i) mean protein numbers are large and (ii) fluctuations are small compared to means.  Under the LNA, the steady state solution to the master equation is a Gaussian distribution: the first-order terms recover the deterministic system and thus provide the means, and the second-order terms yield an equation for the covariance matrix.  Comprehensive discussions of the linear noise approximation can be found in \citep{vanKampen, Elf, Paulsson}.

Under the LNA the steady state distribution over each species' protein number is a Gaussian expansion around the deterministic mean given by the steady state of Eqns.\ \ref{eq:dyn1}-\ref{eq:dyn3}.
The covariance matrix $\Xi$ is determined from model parameters by solving the Lyapunov equation
\beq
\label{eq:lyap}
J\Xi+\Xi J^T+D=0,
\eeq
where $J$ is the Jacobian of the system (Eqn.\ \ref{eq:Jac}) and
\beq
D = \begin{pmatrix}
	R_A(\alpha+A) & 0 & 0 \\
	0 & R_B(\beta+B) & 0 \\
	0 & 0 & R_G(\gamma+G) \\
\end{pmatrix}
\eeq
is an effective diffusion matrix.  We solve Eqn.\ \ref{eq:lyap} numerically using MATLAB's {\tt lyap} function.

The deterministic steady state and the covariance matrix are computed four times (once in each of the chemical input states $c$); the lower-right terms of each provide the mean and variance, respectively, of each (Gaussian) output distribution $p(G|c)$.  The input-output mutual information is computed directly from the distributions $p(G|c)$, as described next.

\subsection{Information theory}

In this study we make use of a central measure from information theory: {\it mutual information} (MI).  MI is a fundamental measure of the strength a relationship between two random variables.  More precisely, it measures the reduction in entropy of one variable given measurement of the other.  MI captures correlation between two random variables even when a relationship exists that is nonlinear (unlike, e.g., the correlation coefficient) or non-monotonic (unlike, e.g., Spearman's rho).  It has found wide use in the study of biological networks both as a statistical measure \citep{Margolin, Ziv2} and as a measure of functionality in the presence of noise \citep{Tkacik2, Mehta, WalczakII}.  For a comprehensive review of information theory we refer the reader to \citep{Cover}.

In this study we employ MI in two separate contexts: (i) as a measure of network functionality and (ii) as a measure of statistical correlation.  In the first context, we optimize the MI between the chemical input state and the concentration of the output protein.  In the second context, we compute, for a given topological feature (e.g.\ number of up-regulations), the MI between the value that the feature takes in a network and the function that the network performs.  We here describe these computations in turn.

MI is the average of the log of a ratio.  The average is over the joint probability distribution between two random variables, and the ratio is that of the joint to the product of the marginal distributions.  For a discrete variable, i.e.\ the chemical input state $c$, and a continuous variable, i.e.\ the output protein concentration $G$, MI takes the form
\beq
I[p(c,G)] = \sum_c \int dG\, p(c,G) \log_2 \frac{p(c,G)}{p(c)p(G)},
\eeq
where the log is taken in base $2$ to give $I$ in bits.  Noting that $p(c,G) = p(G|c)p(c)$ and $p(G) = \sum_c p(c,G)$ allows one to write the MI as
\beq
\label{eq:Icond}
I[p(c,G)] = \sum_c \int dG\, p(G|c) p(c) \log_2 \frac{p(G|c)}{\sum_{c'} p(G|c') p(c')},
\eeq
i.e.\ entirely in terms of the conditional output distributions $p(G|c)$, obtained via the linear noise approximation (see above), and the input distribution $p(c)$, which we take as $p(c) = 1/4$ for equally likely chemical input states.  Eqn.\ \ref{eq:Icond} is integrated numerically during the optimization using MATLAB's {\tt quad}.

In the context of correlating, for a topological feature $\mu$, the feature value $v_\mu$ with the observed function $r$, since both are categorical (and therefore discrete) variables, MI takes the form
\beq
I[p(v_\mu,r)] = \sum_{v_\mu,r} p(v_\mu,r) \log \frac{p(v_\mu,r)}{p(v_\mu)p(r)}.
\eeq
Here, the joint distribution $p(v_\mu,r)$ is computed from the optimization data according to Eqn. 2, and the marginal distributions are trivially obtained as $p(v_\mu) = \sum_r p(v_\mu,r)$ and $p(r) = \sum_{v_\mu} p(v_\mu,r)$.

Additionally we note that MI is again used in the context of statistical correlation to compute a feature adjacency matrix (see Sec.\ \ref{sec:MDS}).  That is, the MI between the values of a feature $\mu$ and a feature $\nu$ is given by
\beq
I[p(v_\mu,v_\nu)] = \sum_{v_\mu,v_\nu} p(v_\mu,v_\nu) \log \frac{p(v_\mu,v_\nu)}{p(v_\mu)p(v_\nu)},
\eeq
where $p(v_\mu,v_\nu)$ is computed directly from the network set.

\subsection{Optimization}

Optimization of input-output information $I[p(G,c)]$ (Eqn.\ \ref{eq:Icond}) over model parameters is done numerically using MATLAB's {\tt fminsearch}.  Each optimization is performed at constrained average protein number $N \equiv (A+B+G)/3$
and average timescale separation $T \equiv [(R_A+R_B)/2]/R_G$ by maximizing the quantity
\beq
L \equiv I[p(G,c)] - \eta N - \kappa T
\eeq
for various values of the Lagrange multipliers $\eta$ and $\kappa$ ($R_G$ is fixed).
The optimization is initialized by sampling uniform-randomly in the logs of the parameters; bounds from which initial parameters are drawn are given in Table \ref{boundtable}.

\begin{table}
\begin{tabular}{|l |l |}
\hline
Parameter & Bounds \\
\hline
Promoter strengths, $s$ & $10^{-4}-10^{6}$ \\
Interaction strengths, $w>1$ (up-regulation) & $1.05-10^2$ \\
Interaction strengths, $w<1$ (down-regulation) & $10^{-2}-0.95$ \\
Binding constants, $K$ & $10^{-1}-10^{2}$ \\
Leakiness, $q$ & $10^{-10}-10^{-2}$ \\
Scaling factors, $x>1$, $y>1$ & $1.1-10^4$ \\
Degradation rates, $R_A$, $R_B$ & $10^{-7}-10^{0}$ \\
Degradation rate, $R_G$ (fixed) & $4\cdot10^{-4}$ \\
\hline
\end{tabular}
\caption{Bounds from which parameters are drawn to initialize optimization.}
\label{boundtable}
\end{table}

\subsection{Robustness of $\rho$ ranking}

The correlation between topological feature value $v_\mu$ and network function $r$ is measured for each feature $\mu$ using a normalized mutual information $\rho_\mu$.  This measure is a function of the joint distribution $p(v_\mu,r)$, whose computation (Eqn.\ 2) depends on two distributions which we take to be uniform: (i) $p(n)$, the probability of observing each network $n$, and (ii) $p(\th|n)$, the probability of observing each optimally functional point $\th$ in the parameter space of network $n$.  Here we show that the ranking of the $\rho_\mu$ is robust to perturbations in the uniformity of each of these distributions.

\subsubsection{Perturbing $p(n)$}

The uniformity of $p(n)$ is perturbed by artificially setting $p(n) \propto (u_n)^\epsilon$, where $u_n$ is a vector of random numbers and $\epsilon$ tunes the entropy of the distribution $H[p(n)]$.  That is, $\epsilon=0$ recovers the maximum-entropy (uniform) distribution, while $\epsilon \rightarrow \infty$ produces the zero-entropy solution $p(n) \rightarrow \delta(n,{\rm argmax} \, u_n)$ (where $\delta$ is the Kronecker delta).  Fig.\ \ref{fig:robust}A plots the $\rho_\mu$ as a function of the entropy $H[p(n)]$.
As seen in Fig.\ \ref{fig:robust}A, the ranking of the top $4$ features is preserved under $\sim$$15\%$ perturbations in the entropy, and that of the top $3$ features is preserved under $\sim$$30\%$ perturbations.  This demonstrates that the feature ranking is considerably robust to perturbations in the uniformity of $p(n)$.

\subsubsection{Perturbing $p(\th|n)$}

\begin{figure}
\begin{center}
\includegraphics[scale=.3]{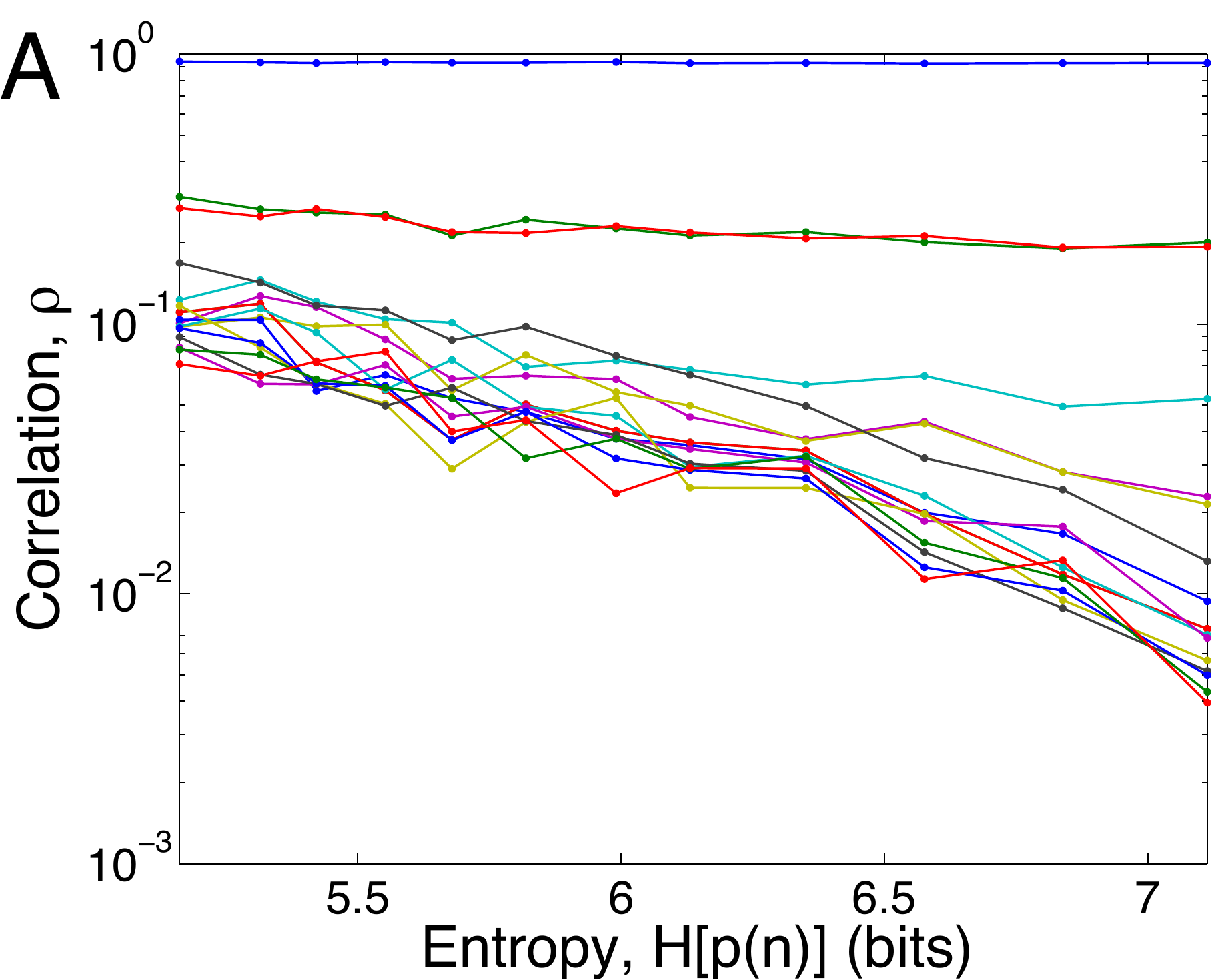}
\includegraphics[scale=.3]{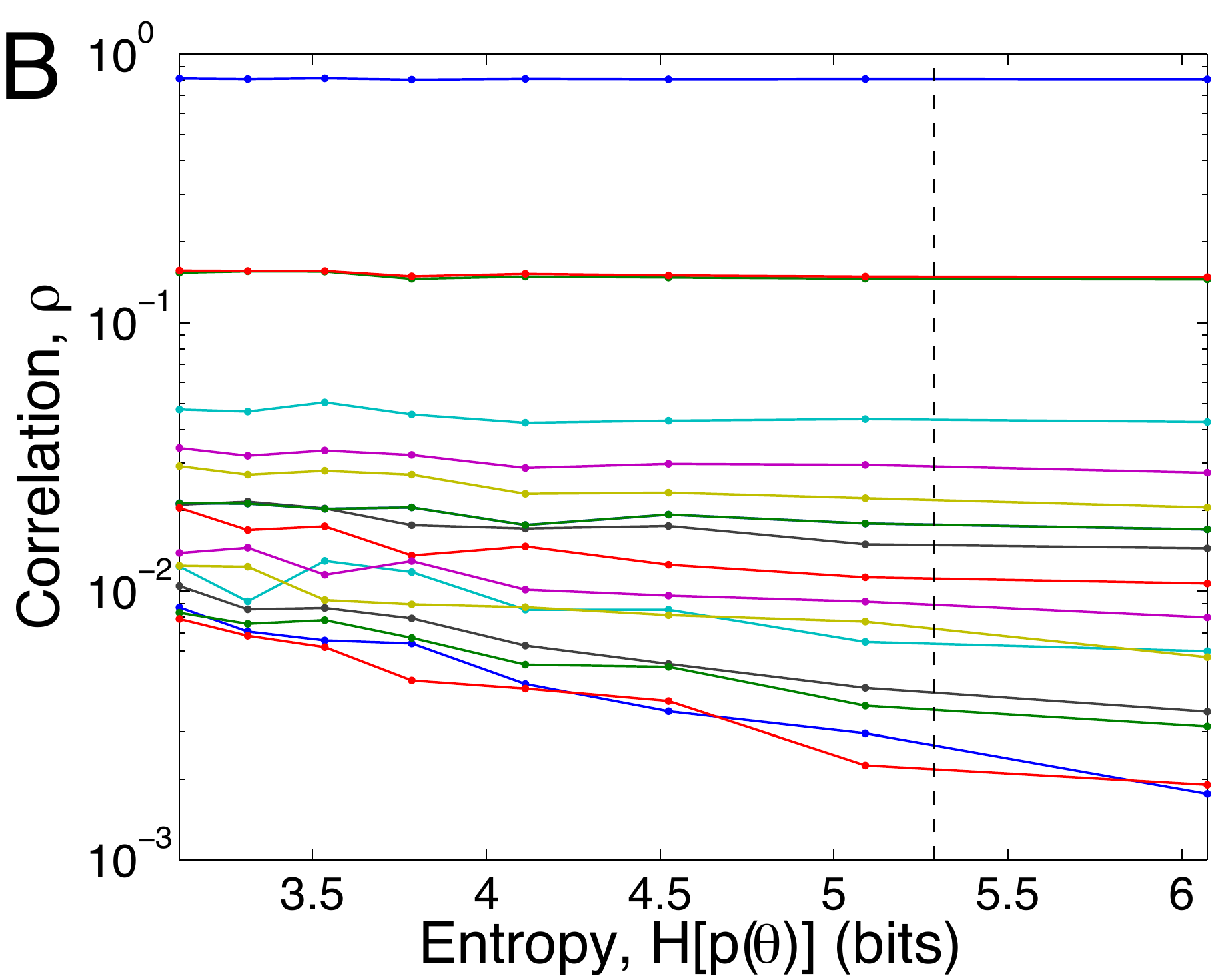}
\caption{Values of the correlation measure $\rho$ for each of the $17$ features as a function of the entropies (A) $H[p(n)]$ and (B) $H[p(\th|n)]$.  Each point represents the average of $8$ trials.  In B, the dashed vertical line shows the entropy under the assumption that $p(\th_0|n)$, not $p(\th|n)$, is uniform.}
\label{fig:robust}
\end{center}
\end{figure}

The uniformity of $p(\th|n)$ for each network $n$ is perturbed via the same procedure described in the previous section.  Fig.\ \ref{fig:robust}B plots the $\rho_\mu$ as a function of the entropy of $p(\th) = \sum_n p(\th|n) p(n)$, where $p(n)$ here is uniform.  As seen in Fig.\ \ref{fig:robust}B, the ranking of the top $7$ features is preserved under $\sim$$40\%$ perturbations in the entropy of $p(\th)$, indicating that the feature ranking is very robust to perturbations in $p(\th|n)$.

In this case we also have an independent entropy scale, given by the fact that we may decompose $p(\th|n)$ as
\beq
p(\th|n) = \sum_{\th_0} p(\th|\th_0,n)p(\th_0|n),
\eeq
where $\th_0$ is the parameter setting that initializes an optimization and $p(\th|\th_0,n)$ is determined by the optimization itself.  If we assume uniformity of $p(\th_0|n)$, instead of $p(\th|n)$, then $p(\th|n)$ is computable from the numbers of times the optimization converges repeatedly on each local optimum $\th$.  The entropy in this case is $13\%$ different from that of the uniform distribution, and the ranking of $\rho$ is almost entirely unchanged (Fig.\ \ref{fig:robust}B).  Fig.\ \ref{fig:robust}B demonstrates that the results are not sensitive to whether one takes the distribution of initial parameters or the distribution of optimal parameters to be uniform.

\subsection{Non-redundant features}\label{sec:MDS}

\begin{figure}
\begin{center}
\includegraphics[scale=.3]{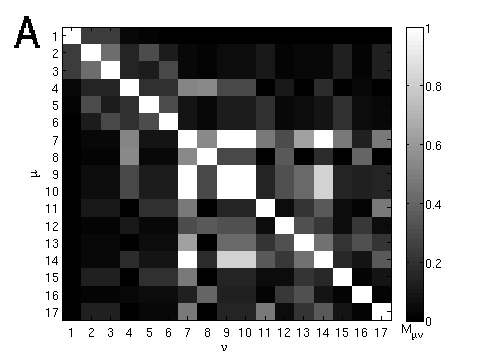}
\includegraphics[scale=.3]{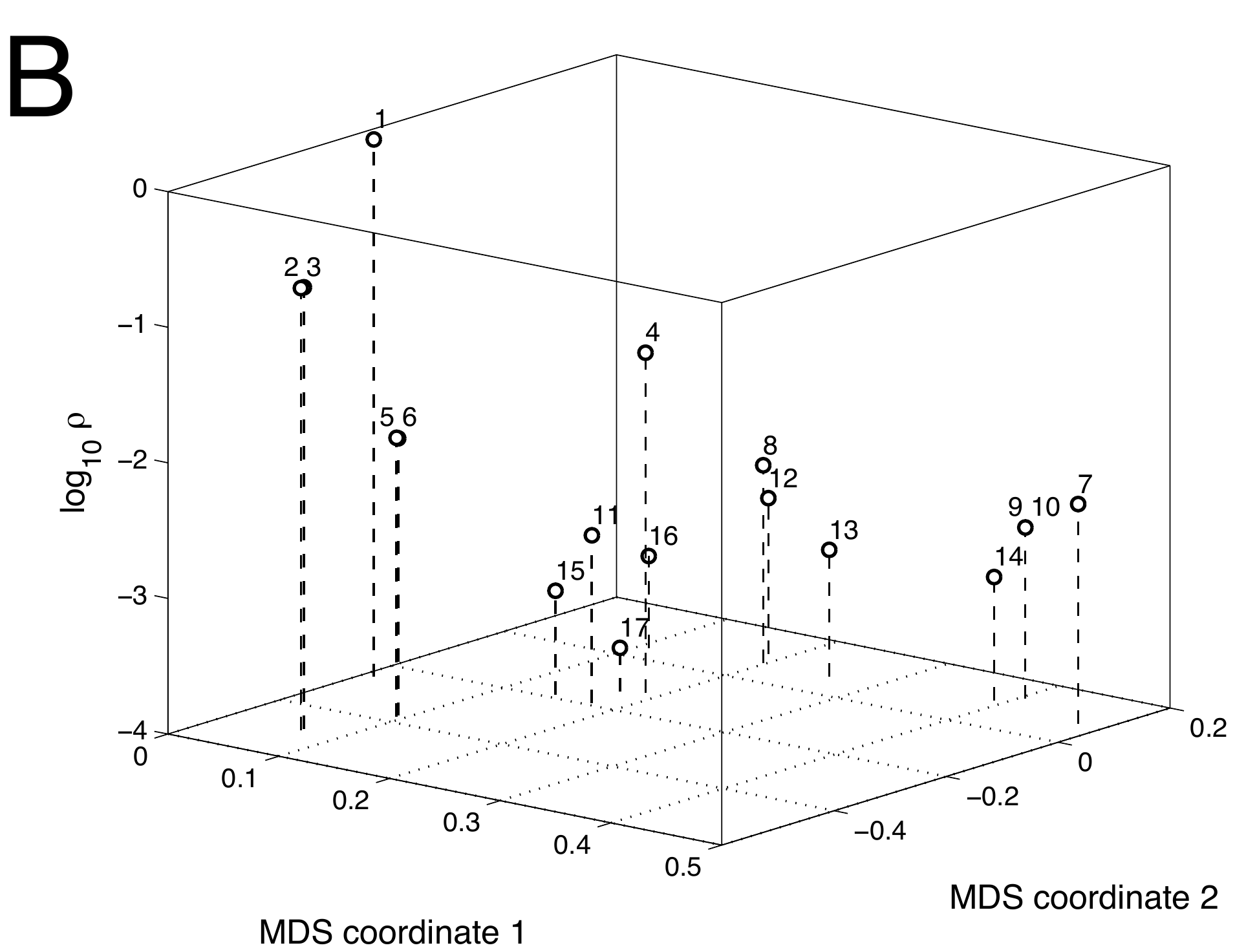}
\caption{Identifying non-redundant topological features.  (A) Feature adjacency matrix (Eqn.\ \ref{eq:featureadj}).  (B) Features plotted according to correlation measure $\rho$ and the coordinates of a two-dimensional scaling based on the adjacency matrix in A.  Features are numbered as in Table \ref{rhotable}.}
\label{fig:mds}
\end{center}
\end{figure}

To interpret which features are associated with which sets of
realizable functions, 
it is useful to group nearly identical features together
and use only the feature which is most informative about function
(highest in $\rho$)
as the exemplar among each group.
To quantify redundancy among features, we compute the MI between each pair of features and normalize by the minimum entropy to produce a weighted adjacency matrix
\beq
\label{eq:featureadj}
M_{\mu\nu} = \frac{I[p(v_\mu,v_\nu)]}{\min\{H[p(v_\mu)],H[p(v_\nu)]\}}
\eeq
(Fig.\ \ref{fig:mds}A).  We then use the adjacency matrix as the basis for multidimensional scaling, as described below.

Multidimensional scaling (MDS) describes a class of techniques used to visualize proximities among data in a low-dimensional space.  One of the most common techniques (also called principal components analysis when applied to a correlation matrix) is to use as the low-dimensional coordinates the eigenvectors of the adjacency matrix corresponding to the largest-magnitude eigenvalues.  Data points with high mutual proximity then tend to be grouped together along these coordinates.  Fig.\ \ref{fig:mds}B and Fig.\ 5 show the application of this technique in two and one dimensions, respectively, to the adjacency matrix in Fig.\ \ref{fig:mds}A, revealing groups of similar features.  Plotting the feature-function correlation measure $\rho$ along the vertical axis in each case makes apparent the most informative feature in each group (i.e.\ feature 1 in one group and feature 4 in a second group).

\subsection{Direct functionality: validation of known analytic result}

In previous work \cite{Mugler1} we show analytically that networks in which each species is regulated by at most one other species perform only ``direct'' functions, in which the sign of the effect of an input species on an output species depends only on the direct path from input to output, even when there is feedback.  This analytic result is here validated by our statistical approach.

\begin{figure}
\begin{center}
\includegraphics[scale=.4]{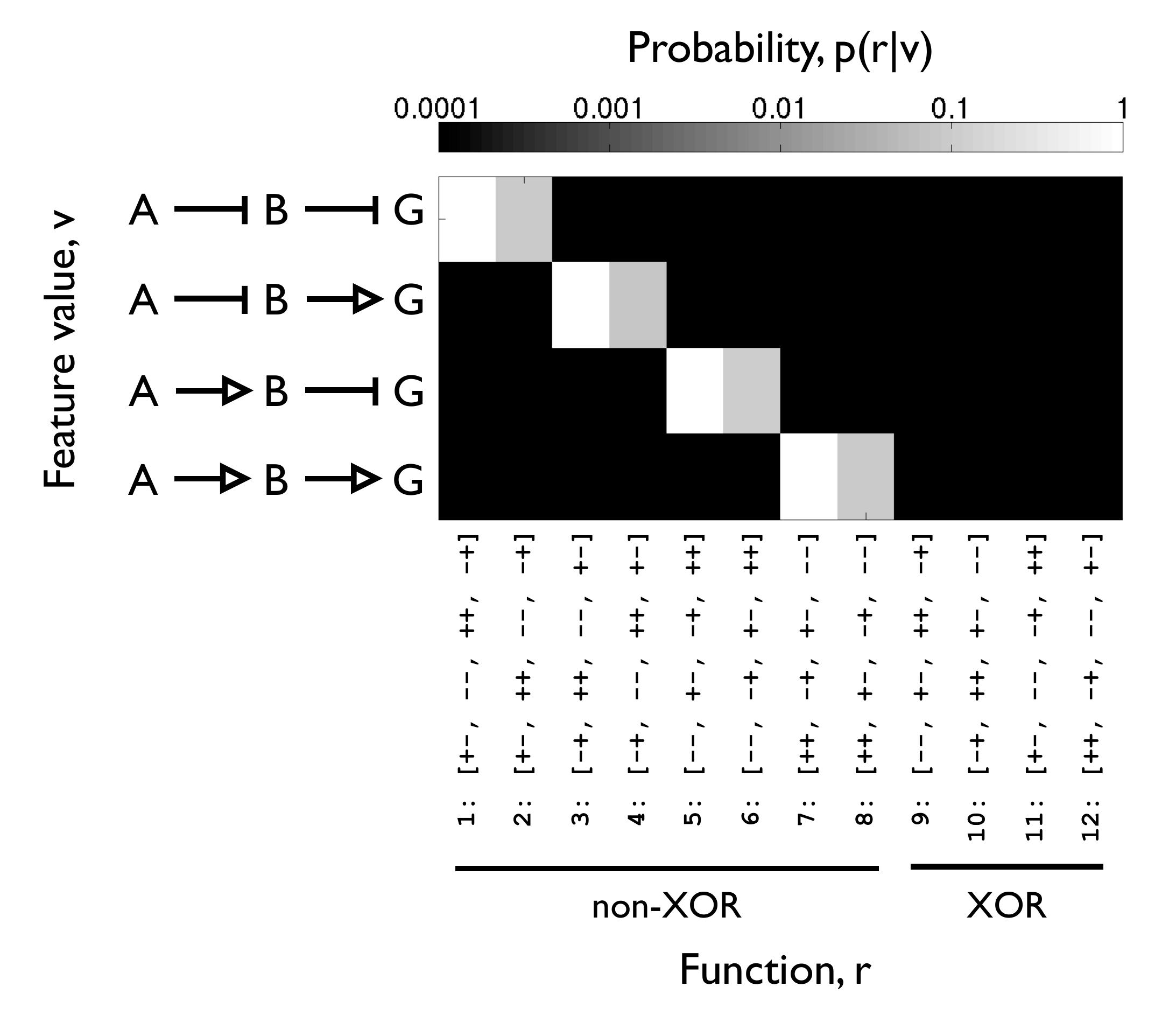}
\caption{Conditional distribution showing the probability of a particular input-output function given the value of the topological feature `signs (up- or down-regulating) of the forward regulatory edges,' using only data from networks in which each species is singly regulated (Fig.\ \ref{fig:topos}A-B).  Only direct functions are performed, as defined in the text.}
\label{fig:valid}
\end{center}
\end{figure}

In the context of our setup (see Fig.\ 1A), the direct paths from the inputs (the chemical inhibitors labeled by $x$ and $y$) to the output (the fluorescent protein $G$) involve only the forward regulatory edges $A \rightarrow B$ and $B \rightarrow G$.  Therefore considering only those networks in which each species is singly regulated (Fig.\ \ref{fig:topos}A-B), the analytic result predicts that the signs of the forward edges uniquely determine the type of function performed.  Plotting the conditional distribution $p(r|v)$ for the feature `signs of the forward edges' using only data from these networks confirms that this is indeed the case (Fig.\ \ref{fig:valid}).  Accordingly the correlation for this distribution is $\rho = 1$, the maximum possible value.

The functions performed at each of the feature values in Fig.\ \ref{fig:valid} can be understood intuitively.  For example, in networks with the last feature value $A \rightarrowtriangle B \rightarrowtriangle G$, inhibition of $A$ and of $B$ will both reduce the expression of $G$, such that the state in which both small molecules are present ($++$) produces the lowest-ranked output, and conversely, the state in which both small molecules are absent ($--$) produces the highest-ranked output; one may verify by inspection that functions $7$ and $8$ are the two that satisfy these criteria.  The correspondence of the other function pairs to feature values can be similarly understood in terms of the edge signs.

\subsection{XOR functionality: analysis}

As described in the main text, XOR functions satisfy one or both of two properties:
\beqn
\label{eq:xor1}
{\rm XOR \,\, property \,\, I:} &{\rm sign} \left( dG/dx \right) {\rm \,\, depends \,\, on \,\,} y,& \\
\label{eq:xor2}
{\rm XOR \,\, property \,\, II:} &{\rm sign} \left( dG/dy \right) {\rm \,\, depends \,\, on \,\,} x.&
\eeqn
To analytically understand the observed XOR functionality, we here calculate the derivatives $dG/dx$ and $dG/dy$ from the steady state of the deterministic system (Eqns.\ \ref{eq:dyn1}-\ref{eq:dyn3}).  We further show how the forms of these derivatives support the observations that (i) all XOR functions satisfying property I are performed by networks in which species $B$ is autoregulated, and (ii) no functions satisfying property II are observed.

\subsubsection{Calculating the derivatives}\label{sec:deriv}

The steady state of Eqns.\ \ref{eq:dyn1}-\ref{eq:dyn3},
with all functional dependencies made explicit, is
\beqn
A &=& \alpha[a(A,x),b(B,y)], \\
B &=& \beta[a(A,x),b(B,y)], \\
G &=& \gamma[b(B,y)],
\eeqn
where $a(A,x) = A/x$ and $b(B,y) = B/y$.
The output $G$ depends on the input $x$ only through $b$, and $b$ depends on $x$ only through $B$,
i.e.\
\beq
\label{eq:dGdxtemp}
\d{G}{x} = \d{\gamma}{x} = \d{\gamma}{b}\pd{b}{B}\d{B}{x}.
\eeq
The dependencies of $A$ and $B$ on $x$ are coupled:
\beqn
\d{A}{x} &=& \d{\alpha}{x}
	= \pd{\alpha}{a} \d{a}{x} + \pd{\alpha}{b}\d{b}{x} \\
\label{eq:Ax}
	&=& \pd{\alpha}{a} \left( \pd{a}{A} \d{A}{x} + \pd{a}{x} \right)
	+ \pd{\alpha}{b} \left( \pd{b}{B} \d{B}{x} \right), \\
\d{B}{x} &=& \d{\beta}{x}
	= \pd{\beta}{a} \d{a}{x} + \pd{\beta}{b}\d{b}{x} \\
\label{eq:Bx}
	&=& \pd{\beta}{a} \left( \pd{a}{A} \d{A}{x} + \pd{a}{x} \right)
	+ \pd{\beta}{b} \left( \pd{b}{B} \d{B}{x} \right).
\eeqn
Eqns.\ \ref{eq:Ax} and \ref{eq:Bx} form an algebraic system of equations in the variables $dA/dx$ and $dB/dx$, whose solution is
\beqn
\label{eq:dAdx}
\d{A}{x} &=& \frac{1}{-\Delta} \left[ \pd{\alpha}{a} \pd{a}{x} \left( 1 - \pd{\beta}{b} \pd{b}{B} \right)
	\right. \nonumber\\
	&& \left. + \pd{\alpha}{b} \pd{b}{B} \pd{\beta}{a} \pd{a}{x} \right], \\
\label{eq:dBdx}
\d{B}{x} &=& \frac{1}{-\Delta} \pd{\beta}{a} \pd{a}{x},
\eeqn
where
\beqn
\Delta &=& \left( \pd{\alpha}{b} \pd{b}{B} \right) \left( \pd{\beta}{a} \pd{a}{A} \right) \nonumber\\
	&& - \left( \pd{\alpha}{a} \pd{a}{A} - 1 \right) \left( \pd{\beta}{b} \pd{b}{B} - 1 \right) \\
\label{eq:Delta}
&=& \pd{\alpha}{B} \pd{\beta}{A}
	- \left( \pd{\alpha}{A} - 1 \right) \left( \pd{\beta}{B} - 1 \right)
\eeqn
is the determinant of the Jacobian of the dynamical system and is always negative at stable fixed points (Eqn.\ \ref{eq:det}).
Substituting Eqn.\ \ref{eq:dBdx} into Eqn.\ \ref{eq:dGdxtemp} and using $\partial \gamma / \partial B = (\partial \gamma / \partial b)(\partial b / \partial B)$ yields Eqn.\ 4 of the main text,
\beq
\label{eq:dGdxapp}
\frac{dG}{dx} = \frac{1}{-\Delta}\frac{\partial a}{\partial x}
	\frac{\partial \beta}{\partial a}\frac{\partial \gamma}{\partial B}.
\eeq

The output $G$ depends on the input $y$ through $b$, which depends on $y$ either indirectly through $B$ or directly, i.e.\
\beq
\label{eq:dGdytemp}
\d{G}{y} = \d{\gamma}{y} = \d{\gamma}{b}\d{b}{y}
	= \d{\gamma}{b} \left( \pd{b}{B} \d{B}{y} +  \pd{b}{y} \right).
\eeq
As on $x$, the dependencies of $A$ and $B$ on $y$ are coupled:
\beqn
\d{A}{y} &=& \d{\alpha}{y}
	= \pd{\alpha}{a} \d{a}{y} + \pd{\alpha}{b}\d{b}{y} \\
\label{eq:Ay}
	&=& \pd{\alpha}{a} \left( \pd{a}{A} \d{A}{y} \right)
	+ \pd{\alpha}{b} \left( \pd{b}{B} \d{B}{y} + \pd{b}{y} \right), \\
\d{B}{y} &=& \d{\beta}{y}
	= \pd{\beta}{a} \d{a}{y} + \pd{\beta}{b}\d{b}{y} \\
\label{eq:By}
	&=& \pd{\beta}{a} \left( \pd{a}{A} \d{A}{y} \right)
	+ \pd{\beta}{b} \left( \pd{b}{B} \d{B}{y} + \pd{b}{y} \right).
\eeqn
Eqns.\ \ref{eq:Ay} and \ref{eq:By} can be solved to yield
\beqn
\label{eq:dAdy}
\d{A}{y} &=& \frac{1}{-\Delta} \pd{\alpha}{b} \pd{b}{y}, \\
\label{eq:dBdy}
\d{B}{y} &=& \frac{1}{-\Delta} \left[ \pd{\beta}{b} \pd{b}{y} \left( 1 - \pd{\alpha}{a} \pd{a}{A} \right)
	\right. \nonumber\\
	&& \left. + \pd{\beta}{a} \pd{a}{A} \pd{\alpha}{b} \pd{b}{y} \right],
\eeqn
where $\Delta$ is the determinant as in Eqn.\ \ref{eq:Delta}.  Substituting Eqn.\ \ref{eq:dBdy} into Eqn.\ \ref{eq:dGdytemp} and simplifying gives Eqn.\ 5 of the main text,
\beq
\label{eq:dGdy}
\frac{dG}{dy} = \frac{1}{-\Delta} \left( 1 - \frac{\partial \alpha}{\partial A} \right)
	\frac{\partial b}{\partial y}\frac{d \gamma}{d b},
\eeq
where $\partial \alpha / \partial A = (\partial \alpha / \partial a)(\partial a / \partial A)$.

\subsubsection{Type-I XOR functions require autoregulation of $B$}\label{sec:prop1}

Type-I XOR functionality (Eqn.\ \ref{eq:xor1}) requires the sign of the derivative $dG/dx$ (Eqn.\ \ref{eq:dGdxapp}) to depend on $y$.  Here we go through each term in Eqn.\ \ref{eq:dGdxapp} and conclude that only the third can change sign.  The first term in Eqn.\ \ref{eq:dGdxapp}, $1/(-\Delta)$, is always positive because the determinant $\Delta$ is equal to the product of the three eigenvalues of the Jacobian of the deterministic system, which at a stable fixed point are all negative (Eqn.\ \ref{eq:det}).  The second term in Eqn.\ \ref{eq:dGdxapp} is $\partial a / \partial x = -A/x^2$, which is always negative, corresponding to the inhibitory effect of the small molecule $x$ on transcription factor $A$.  The fourth term in Eqn.\ \ref{eq:dGdxapp} is $\partial \gamma / \partial B = (\partial \gamma / \partial b)(\partial b / \partial B)$.  The factor $\partial b / \partial B = 1/y$ is always positive, and the factor $\partial \gamma / \partial b$ is of unique sign because $\gamma(b)$ is monotonic, as shown below.  This leaves only the third term, $\partial \beta / \partial a$, which can change sign if and only if $B$ is autoregulated, as discussed below.

Under our model a regulation function with only one argument is monotonic, which is consistent with the interpretation of the edge being either up- or down-regulating.  For example, the regulation function corresponding to the edge $B \rightarrow G$ in all networks is (see Eqns.\ \ref{eq:gamma}-\ref{eq:prob})
\beq
\gamma(b) = \frac{s_G}{R_G} \frac{Z^G_{\rm on}}{Z^G_{\rm on} + Z^G_{\rm off}},
\eeq
where $Z_{\rm on}^G$ and $Z_{\rm off}^G$ are given by, e.g., Eqns.\ \ref{eq:ZGon}-\ref{eq:ZGoff}.
The derivative of this function with respect to its argument is
\beq
\label{eq:dform}
\d{\gamma}{b} = \frac{s_G}{R_GZ_G^2}
	\left( \d{Z^G_{\rm on}}{b} Z^G_{\rm off} - Z^G_{\rm on} \d{Z^G_{\rm off}}{b} \right),
\eeq
where $Z_G = Z^G_{\rm on} + Z^G_{\rm off}$.  Upon differentiating and inserting
Eqns.\ \ref{eq:ZGon} and \ref{eq:ZGoff},
all dependence on $b$ inside the parentheses cancels, leaving
\beq
\label{eq:mono}
\d{\gamma}{b} = \frac{s_Gq}{R_GK_2^GZ_G^2}(w_{02}^G-1).
\eeq
Eqn.\ \ref{eq:mono} confirms that $\gamma(b)$ is monotonic, with $w_{02}^G > 1$ corresponding to up-regulation, and $w_{02}^G < 1$ corresponding to down-regulation.

If species $B$ is not autoregulated, then it is only regulated by species $A$; the regulation function then only has one argument, i.e.\ $\beta(a,b) = \beta(a)$, and, as with $\gamma(b)$ above, it is monotonic.  Therefore without autoregulation of $B$, type-I XOR functionality is not possible.  We now compute the derivative $\partial \beta / \partial a$ in the case of two arguments to analytically demonstrate the converse: that with autoregulation of $B$, type-I XOR functionality is possible.

As with Eqn.\ \ref{eq:dform}, the partial derivative of $\beta(a,b)$ with respect to $a$ takes the form
\beq
\pd{\beta}{a} = \frac{s_B}{R_BZ_B^2}
	\left( \pd{Z^B_{\rm on}}{a} Z^B_{\rm off} - Z^B_{\rm on} \pd{Z^B_{\rm off}}{a} \right),
\eeq
where $Z_B = Z^B_{\rm on} + Z^B_{\rm off}$, $Z_{\rm on}^B$ (with $B$ autoregulated) is given by, e.g., Eqn.\ \ref{eq:ZBadd} for additive interaction and Eqn.\ \ref{eq:ZBmult} for multiplicative interaction, and $Z_{\rm off}^B$ is given by, e.g., Eqn.\ \ref{eq:ZBoff2}.  Upon differentiating and inserting the expressions for $Z_{\rm on}^B$ and $Z_{\rm off}^B$, all dependence on $a$ inside the parentheses cancels (for both additive and multiplicative interaction), leaving
\beq
\label{eq:quad}
\pd{\beta}{a} = \frac{s_Bq}{R_BK_1^BZ_B^2}
	\left( C_2 b^2 + C_1 b + C_0 \right),
\eeq
where for additive interaction,
\beqn
C_0 &=& w_{01}^B - 1, \\
C_1 &=& \frac{1}{K_2^B} \left[ w_{01}^B - w_{02}^B - w_{12}^B +
	\left( w_{01}^B + w_{02}^B \right) w_{12}^B \right], \quad \\
C_2 &=& \left( \frac{1}{K_2^B} \right)^2 w_{01}^B w_{12}^B,
\eeqn
and for multiplicative interaction,
\beqn
C_0 &=& w_{01}^B - 1, \\
C_1 &=&  \frac{1}{K_2^B} \left[ w_{01}^B - w_{02}^B - w_{12}^B + 
	w_{01}^B w_{02}^B w_{12}^B \right], \\
C_2 &=& \left( \frac{1}{K_2^B} \right)^2 \left( w_{01}^B - 1 \right) w_{02}^B w_{12}^B.
\eeqn
Eqn.\ \ref{eq:quad} is the product of a positive term and a quadratic function of $b = B/y$.  It is straightforward to demonstrate in both the additive and multiplicative cases (e.g.\ by sampling numerically) that for positive $w_{01}^B$, $w_{02}^B$, and $w_{12}^B$ the quadratic function can have positive, negative, or complex roots.  When at least one root is positive, the sign of $\partial \beta / \partial a$ changes at positive $B/y$, i.e.\ the sign can depend on $y$.  Since $dG/dx$ is proportional to $\partial \beta / \partial a$ (Eqn.\ \ref{eq:dGdxapp}), this enables type-I XOR functionality. This analysis suggests inspection of the parameters themselves obtained via optimization; doing so, we observe that the vast majority of observed XOR functions results from optimal parameter values for which there exists a positive root in the range $0 < B/y < $ $\sim$$100$, which is precisely the range of protein numbers in which our optimal solutions lie.

To summarize, nonmonotonicity in the regulation of species $B$, which can occur only when $B$ is autoregulated, produces the observed XOR functions.

\subsubsection{Type-II XOR functions are not observed}\label{sec:prop2}

Type-II XOR functionality (Eqn.\ \ref{eq:xor2}) requires the sign of the derivative $dG/dy$ (Eqn.\ \ref{eq:dGdy}) to depend on $x$.  Three of the terms in Eqn.\ \ref{eq:dGdy} are of unique sign: the terms $1/(-\Delta)$ and $d\gamma/db$ are positive and of unique sign respectively, as discussed in the previous section; and the term $\partial b / \partial y = -B/y^2$ is always negative, corresponding to the inhibitory effect of the small molecule $y$ on transcription factor $B$.  This leaves only the term $(1-\partial\alpha/\partial A)$, which, as discussed below, for four of the network topologies is provably positive at stable fixed points, and for the other two network topologies is observed to be positive for all optimal solutions.

For topologies B and E (Fig.\ \ref{fig:topos}), in which $\partial \alpha /\partial A = 0$, the term $(1-\partial\alpha/\partial A) = 1$ is clearly positive.  For topologies A and C, in which $\partial \alpha /\partial B = 0$, the first eigenvalue of the Jacobian (Eqn.\ \ref{eq:eig}) reduces to $\lambda_1 = \partial \alpha /\partial A - 1$; since this must be negative for stability, the term $(1-\partial\alpha/\partial A)$ is always positive for these topologies as well.  This leaves only networks with topology D or F.

In networks with topology D or F, It is unclear whether type-II XOR functions are analytically forbidden or simply exceedingly improbable for optimally informative parameters.  Some analytic constraints can be obtained by the facts that since the real parts of $\lambda_1$ and $\lambda_2$ (Eqn.\ \ref{eq:eig}) are negative at stable fixed points, their sum must be negative and their product must be positive, i.e.\
\beqn
\label{eq:con1}
0 &<& -(\lambda_1 + \lambda_2) = \left( 1 - \pd{\alpha}{A} \right) + \left( 1 - \pd{\beta}{B} \right), \\
\label{eq:con2}
0 &<& \lambda_1 \lambda_2 = \left( 1 - \pd{\alpha}{A} \right) \left( 1 - \pd{\beta}{B} \right)
	- \pd{\alpha}{B} \pd{\beta}{A}. \quad
\eeqn
If $(1 - \partial \beta / \partial B) < 0$, then Eqn.\ \ref{eq:con1} implies that $(1-\partial\alpha/\partial A) > 0$, forbidding type-II XOR functions.  If on the other hand $(1 - \partial \beta / \partial B) > 0$, Eqn.\ \ref{eq:con1}  does not restrict the sign of $(1-\partial\alpha/\partial A)$, but Eqn.\ \ref{eq:con2} implies
\beq
\left( 1 - \pd{\alpha}{A} \right) > \frac{(\partial \alpha / \partial B)(\partial \beta / \partial A)}{1 - \partial \beta / \partial B}.
\eeq
Although in this last case it is not known whether the right-hand side is constrained to be positive, empirically the term $(1-\partial\alpha/\partial A) > 0$ is observed to be positive for all optimal solutions, even over wide variations in the orders of magnitude of each of the optimal parameters across solutions.

\begin{acknowledgments}
The authors thank Nicolas Buchler and William Bialek for useful conversations.  AM was supported by NSF grant DGE-0742450; CHW was supported by NIH grants 1U54CA121852-01A1 and 5PN2EY016586-03.
\end{acknowledgments}

\end{document}